\documentclass{elsart3p}
\usepackage{graphicx,amssymb}
\usepackage{amsmath}
\usepackage[usenames]{color}

\journal{Physica D}

\input{tcilatex}
\begin{document}

\begin{frontmatter}

\title{Gap solitons in a model of a superfluid fermion gas in optical lattices}

\author{Sadhan
K. Adhikari$^{1}$,
Boris A. Malomed$^{1,2}$}
\address{$^1$Instituto de F\'{\i}sica Te\'{o}rica, UNESP -- S\~{a}o
Paulo State
University, 01.405-900 S\~{a}o Paulo, S\~{a}o Paulo, Brazil\\
$^2$Department of Physical Electronics, School of Electrical
Engineering,
Tel Aviv University, Tel Aviv 69978, Israel}

\begin{abstract} We consider a dynamical model for a Fermi gas in the
Bardeen-Cooper-Schrieffer (BCS) superfluid state, trapped in a
combination of a 1D or 2D optical lattice (OL) and a tight parabolic
potential acting in the transverse direction(s). The model is based on
an equation for the order parameter (wave function), which is derived
from the energy density for the weakly coupled BCS superfluid. The
equation includes a nonlinear self-repulsive term of power $7/3$, which
accounts for the Fermi pressure. Reducing the equation to the 1D or 2D
form, we construct families of stable 1D and 2D gap solitons (GSs) by
means of numerical simulations, which are guided by the variational
approximation (VA). The GSs are, chiefly, compact objects trapped in a
single cell of the OL potential. In the linear limit, the VA predicts
almost exact positions of narrow Bloch bands that separate the
semi-infinite and first finite gaps, as well as the first and second
finite ones. Families of stable even and odd bound states of 1D GSs are
constructed too. We also demonstrate that the GS can be dragged without
much distortion by an OL moving at a moderate velocity ($\sim $ 1 mm/s,
in physical units). The predicted GSs contain $\sim 10^{3}-10^{4}$ and
$\sim 10^{3}$ atoms per 1D and 2D settings, respectively. \end{abstract}

\end{frontmatter}

\section{Introduction}

Studies of matter-wave patterns in degenerate Bose \cite{books} and Fermi
\cite{Fermi} gases at ultra-low temperatures have drawn a great deal of
attention in the last decade. The theoretical description of patterns formed
in a Bose-Einstein condensate (BEC) relies upon the Gross-Pitaevskii
equation (GPE), which produces a remarkably accurate description of various
states, including solitons \cite{BECsolitons}. Experimentally, bright
matter-wave solitons were created in BEC of $^{7}$Li \cite{Li-soliton} and $%
^{85}$Rb atoms \cite{Rb-soliton}. In either case, the interaction between
the atoms was switched from repulsion to attraction by means of the magnetic
field applied near the respective Feshbach resonance (FR) \cite%
{Feshbach,bffesh}. In BEC with repulsion between atoms, loaded in a periodic
optical-lattice (OL) potential, which can be created as a standing wave by
counter-propagating laser beams illuminating the condensate, localized
states can be formed in the form of \textit{gap solitons} (GSs). This was
predicted taking into consideration the possibility of having a negative
effective mass in parts of the bandgap spectrum induced by the periodic
potential \cite{GSprediction,Sakaguchi}. GSs are supported by the interplay
of the negative effective mass and repulsive nonlinearity. They are stable
against small perturbations \cite{Pelinovsky}. The prediction was followed
by the creation of a GS formed by $\simeq 250$ atoms of $^{87}$Rb in a
nearly one-dimensional (1D), i.e., \textquotedblleft cigar-shaped",
cross-beam optical trap, combined with the longitudinal OL potential \cite%
{Markus,Markus-review}. In the experiment, the gas was subjected to
acceleration, in order to push atoms into a state with a negative effective
mass. Other theoretically elaborated possibilities for the creation of GSs
rely on the phase-imprinting method \cite{Veronica}, or on temporarily
adding a strong parabolic trap to the OL potential, with the objective to
squeeze the condensate into a small region and then, relaxing the trap, 
to
give the condensate a chance to self-trap into a compact GS state \cite%
{Michal}.

The creation of degenerate Fermi gases (DFG), and observation of the
transition to Bardeen-Cooper-Schrieffer (BCS) superfluidity in them \cite%
{Jin}, suggest to consider possibilities of the creation of solitons in DFGs
and BCS superfluids. Bright solitons were also predicted in Fermi-Bose
mixtures with repulsion \cite{BFsoliton(BBrepBFattr),Sadhan-BFsoliton} or
attraction \cite{Sadhan-BFsoliton} between bosons and strong attraction
between fermions and bosons.

A possibility of the existence of fermionic GSs was mentioned in work \cite%
{Arik}, in relation to a model of a binary boson condensate trapped in a
one- or two-dimensional OL, with repulsion between the two components and
zero intra-species interaction, as the respective system of coupled GPEs may
also be realized as a system of equations for two fermionic wave functions,
and thus the GSs reported in \cite{Arik} may be interpreted as 1D and 2D
\textit{symbiotic} two-component GSs in the binary DFG (term
\textquotedblleft symbiotic" is frequently applied to two-component solitons
in situations when each component in isolation is not able to form a
soliton, while the coupling between them makes it possible \cite{symbio}).
Using more sophisticated models, one-dimensional GSs in the Fermi-Bose
mixture with a small number of Bose atoms were predicted \cite{Kenkre}, and
a possibility to create solitons in a mixed 1D Bose-Fermi superfluid
composed of a relatively small number of atoms was proposed in \cite%
{SadhanLuca}.

In the strictly 1D geometry, the DFG can be mapped into the \textit{%
Tonks-Girardeau gas} of hard-core bosons \cite{Tonks}, which, in a certain
approximation, may be described by the 1D nonlinear Schr\"{o}dinger (NLS)
equation with a repulsive quintic term \cite{further}. In works \cite%
{Fatkhulla}, GSs were predicted in the latter equation which includes the OL
potential, although the solitons reported in those works may be unstable
against small perturbations.

In this work, we aim to predict 1D and 2D GSs in a BCS superfluid, which may
be formed by Fermi atoms with weak attraction between ones with opposite
orientations of the atomic spin. The GSs will be found, chiefly, in the form
of tightly bound states, localized in a single cell of the OL potential. A
rigorous many-body approach to the dynamics of the DFG, based on the system
of quantum-mechanical equations of motion for individual fermions, made it
possible to {demonstrate the formation of fermionic GSs in a Bose-Fermi
mixture} \cite{Mario}. However, such a microscopic approach becomes
unfeasible as the number of fermions increases. A simplified dynamical model
of the BCS superfluid relies on the use of a single wave function for
fermions bound into Cooper pairs. This approach may be compared to that
adopted in the Ginzburg-Landau theory, where the wave function describing
the Fermi superfluid is introduced as a phenomenological complex order
parameter \cite{GL}. Actually, the so developed description of macroscopic
objects, such as solitons, is valid in the \textit{hydrodynamic approximation%
}, i.e., under the condition that the solitons' size is much larger than the
de Broglie wavelength at the Fermi surface, in terms of the underlying
microscopic distribution of the fermion atoms. As a result, the wave
function of the BCS superfluid loaded in the 1D or 2D OL, and confined in
the remaining (transverse) direction(s) by a tight \textquotedblleft
cigar-shaped" or \textquotedblleft pancake" trap, obeys, respectively, the
1D or 2D Schr\"{o}dinger equation, which includes the corresponding OL
potential and a repulsive nonlinear term of power $7/3$ \cite{we}. The
equation is derived from the Lagrangian which includes the energy density
for the weakly-coupled BCS superfluid, as obtained in well-known works \cite%
{Yang3D}.

We construct GS solutions by means of numerical computations, which are
guided by an analytical variational approximation (VA) \cite{VA}; this
approach has produced very accurate results in a model of the Bose-Fermi
superfluid mixture \cite{we2}. The stability of the GSs is established in a
numerical form, through direct simulations of the evolution of perturbed
solitons. We conclude that the VA performs well in describing profiles of
compact \textit{fundamental} GSs sufficiently deep in bandgaps, but it works
poorly near bandgap edges, facing the difficulty in reproducing undulating
tails demonstrated by numerical simulations in that case. In the linear
limit, the approximation very accurately predicts the location of the left
edge of the first finite bandgap.

The paper is organized as follows. The derivation of the 1D and 2D equations
for the wave function is presented in Section II. In Section III, we
construct a family of stable fundamental GS solutions in the first two
finite bandgaps of the 1D model, using the VA based on the Gaussian ansatz
and direct numerical solutions. The GS solutions maintain a tightly-bound
(compact) shape, unless their chemical potential is taken very close to
edges of the bandgaps. In Section III, we also present stable symmetric and
antisymmetric bound states of fundamental GSs, and consider a possibility of
dragging GSs by a moving OL potential. In the framework of the GPE, the
latter issue has drawn considerable attention as a means for transportation
of matter-wave solitons, see work \cite{Panos} and references therein. In
Section IV, we report variational and numerical results for stable GSs in
the 2D equation. In addition to the case of the square-lattice potential, in
Section IV we also find 2D \textit{radial} GSs, supported by an axisymmetric
potential which is periodic along the radius (in BEC trapped in the
axisymmetric OL, radial GSs have been predicted in work \cite{BBBradial}).
Estimates for the number of atoms in the predicted 1D and 2D GSs are given
at the end of Sections III and IV, respectively, the result being $N_{%
\mathrm{1Dsoliton}}\sim 10^{3}-10^{4}$ and $N_{\mathrm{2Dsoliton}}\simeq
2,000$.

For the sake of comparison, in Sections III and IV we additionally display
variational and numerical results for GS families in the ordinary GPE-based
1D and 2D models, which include the corresponding OL potential and,
respectively, the cubic or quintic self-repulsive term. Actually, the
accuracy provided by the VA in the model with the present model, with the
nonlinearity of power $7/3$, is higher than in the models with the cubic or
quintic nonlinear terms.

In addition to the spatially symmetric (even) fundamental GSs, the 1D
equation gives rise to \textit{subfundamental} (SF) solitons, which are
localized odd states originating in the second bandgap. The SF solitons are
also squeezed, essentially, into a single cell of the OL potential, but
unlike the fundamental GSs, they are (weakly) unstable. The SF solitons are
considered in Appendix. In particular, the VA developed for them yields a
physically relevant finding: in the linear limit, it accurately predicts the
position of the narrow Bloch band separating the first and second bandgaps.

\section{Dynamical Equations}

\subsection{The equation in the general form}

We consider a BCS superfluid of Fermi atoms with spin $1/2$ and mass $m$,
assuming weak attraction between fermions with opposite orientations of the
spin. To derive an effective equation for the superfluid order parameter, we
start with the well-known expression for the energy density of the weakly
coupled BCS superfluid in the 3D space \cite{Yang3D,salasnich},
\begin{equation}
\mathcal{E}_{\mathrm{3D}}=(3/5)\rho _{\mathrm{3D}}\varepsilon _{F}+4\pi
a\hbar ^{2}\rho _{\mathrm{3D}}^{2}/(2m)+...,  \label{1A}
\end{equation}%
which is built as an expansion in powers of $k_{F}a$, where $\hbar k_{F}$
and $a$ are the Fermi momentum and scattering length of the weak interaction
between fermions in the BCS limit, $\varepsilon _{F}=\left( \hbar
k_{F}\right) ^{2}/(2m)$ is the Fermi energy, and $\rho _{\mathrm{3D}}$ the
atomic density.
The second term in Eq. (\ref{1A}) is a small correction to the first one due
to the underlying condition, $k_{F}a\ll 1$ (which actually implies that $a$
is much smaller than the de Broglie wavelength at the Fermi surface). Taking
into regard the Cooper pairing of spin-up and spin-down fermions, their
total density is $\rho _{\mathrm{3D}}={2(2\pi )^{-3}}\int_{0}^{k_{F}}4\pi
k^{2}dk\equiv \left( 3\pi ^{2}\right) ^{-1}\left( 2m\varepsilon _{F}/\hbar
^{2}\right) ^{3/2}$. From here, the Fermi energy can be expressed in terms
of the atom density,
\begin{equation}
\varepsilon _{F}=\frac{\hbar ^{2}}{2m}\left( 3\pi ^{2}\rho _{\mathrm{3D}%
}\right) ^{2/3},  \label{mu}
\end{equation}%
hence the energy density in Eq. (\ref{1A}) is cast in the form of
\begin{equation}
\mathcal{E}_{\mathrm{3D}}=\frac{3(3\pi ^{2})^{2/3}\hbar ^{2}}{10m}\rho
_{3D}^{5/3}+\frac{2\pi a\hbar ^{2}}{m}\rho _{\mathrm{3D}}^{2},  \label{1C}
\end{equation}%
which includes the lowest-order correction proportional to scattering length
$a$.

In the framework of the density-functional theory, the BCS superfluid is
described by a complex order parameter (wave function), $\Psi $, such that $%
\rho _{\mathrm{3D}}\equiv |\Psi |^{2}$. If the Fermi energy is much larger
than the depth of the external potential, $V(\mathbf{r})$, the evolution of
the order parameter can be derived from the corresponding Lagrangian density
which includes energy density (\ref{1C}),
\begin{eqnarray}
\mathcal{L} &=&\frac{i\hbar }{2}(\Psi ^{\ast }\Psi _{t}-\Psi \Psi _{t}^{\ast
})-\frac{\hbar ^{2}}{2m_{\mathrm{eff}}}|\nabla \Psi |^{2}-V(\mathbf{r})|\Psi
|^{2}  \notag \\
&&-\frac{3(3\pi ^{2})^{2/3}\hbar ^{2}}{10m}|\Psi |^{10/3}-\frac{2\pi a\hbar
^{2}}{m}|\Psi |^{4},  \label{density}
\end{eqnarray}%
where $m_{\mathrm{eff}}$ is the effective mass of the order-parameter field
in the density-functional theory, which determines the gradient term in the
Lagrangian density. Such a formalism has been used in the description of
Fermi superfluids in various contexts \cite{we3}.

The Euler-Lagrange equation which follows from the full Lagrangian, $L=\int
\mathcal{L}d\mathbf{r}$, where density (\ref{density}) is inserted, takes
the form of the three-dimensional NLS equation with the repulsive nonlinear
term of power $7/3$, and a small correction in the form of the cubic term:%
\begin{equation*}
i\hbar \Psi _{t}=-\frac{\hbar ^{2}}{2m_{\mathrm{eff}}}\left( \partial
_{x}^{2}+\partial _{y}^{2}+\partial _{z}^{2}\right) \Psi
\end{equation*}%
\begin{equation}
+\frac{\hbar ^{2}}{2m}\left[ \left( 3\pi ^{2}\right) ^{2/3}|\Psi
|^{4/3}+8\pi a|\Psi |^{2}\right] \Psi +V(\mathbf{r})\Psi .  \label{threeD1}
\end{equation}%
Since $|\Psi |^{2}$ is defined as the atomic density, Eq. (\ref{threeD1}) is
supplemented by the normalization condition,
\begin{equation}
\int \int \int |\Psi (\mathbf{r},t)|^{2}dxdydz=N,  \label{total}
\end{equation}%
where $N$ is the total number of atoms.

Equation (\ref{threeD1}), which was derived from the local Fermi
distribution, applies to the description of spatially nonuniform patterns in
the hydrodynamic limit, which assumes that the nonuniformity does not
strongly disturb the local distribution. To say it more accurately, the
hydrodynamic approach is valid if the characteristic size of the macroscopic
pattern (actually, the width of the soliton, which is close to the OL
period, $\lambda /2$, see Fig. \ref{fig2} below) is much larger than the de
Broglie wavelength at the Fermi surface, i.e.,
\begin{equation}
\lambda \gg 2\pi /k_{F}.  \label{de Broglie}
\end{equation}%
This condition is similar to that necessary for the validity of the
hydrodynamic approximation in classical rarefied gases: the scale of the
flow must be much larger that the free-path length.

%

\subsection{The one-dimensional equation}

Taking $V(\mathbf{r})$ as the potential of the 3D\ OL, Eq. (\ref{threeD1})
can be used to construct three-dimensional GSs. However, producing
systematic results for 3D solitons is a challenging numerical problem. In
this work, our objective is to reduce the equation to its 1D and 2D forms,
assuming, as mentioned above, that the superfluid is held in a relatively
tight cigar-shaped or pancake-like trap, that correspond to the following
transverse potentials:%
\begin{equation}
V_{\perp }^{\left( \mathrm{cig}\right) }=\frac{1}{2}m\omega _{\perp
}^{2}\left( y^{2}+z^{2}\right) ;V_{\perp }^{\left( \mathrm{panc}\right) }=%
\frac{1}{2}m\omega _{\perp }^{2}z^{2}.  \label{ho}
\end{equation}%
For this reduction to be valid, the Fermi energy must be larger than both
the distance between energy levels in the spectra of harmonic-oscillator
potentials (\ref{ho}), and depth $V_{0}$ of the longitudinal OL potential:%
\begin{equation}
\varepsilon _{F}\gg \hbar \omega _{\perp },V_{0}~.  \label{semiclassical}
\end{equation}%
%
%
Equations (\ref{de Broglie}) and (\ref{semiclassical}) constitute conditions
necessary for the validity of the approach elaborated in the present work.

The next step is reducing Eq. (\ref{threeD1}) to effective 1D and 2D
equations. In the ordinary GPE with the cubic nonlinearity, the reduction of
the 3D equation to its 1D and 2D forms can be performed in different ways,
depending on the particular setting \cite{Luca,CQ,others}. In the simplest
situation, the reduction of the 3D equation to 1D starts with assuming the
factorization of the wave function,
\begin{equation}
\Psi (x,y,z,t)=\Phi (x,t)\exp \left( -\frac{y^{2}+z^{2}}{2na_{\mathrm{ho}%
}^{2}}\right) ,  \label{psiphi}
\end{equation}%
where the transverse harmonic-oscillator length is $a_{\mathrm{ho}}=\sqrt{%
\hbar /\left( m\omega _{\perp }\right) }$, and, in the case of a very tight
confinement, $n=1$. However, due to condition (\ref{semiclassical}), $%
\varepsilon _{F}$ corresponds not to the ground state of the transverse
potential, but rather to an excited state with large quantum number $n$, so
that $\varepsilon _{F}=n\hbar \omega _{\perp }$, and the transverse size of
the domain filled by the Fermi superfluid is $\simeq \sqrt{n}a_{\mathrm{ho}}$%
. Accordingly, the second multiplier in Eq. (\ref{psiphi}) approximates a
lumped superposition of excited states of the transverse harmonic
oscillator, with the corresponding quantum number taking values from $1$ to $%
n$. The self-consistency demands that $\varepsilon _{F}=n\hbar \omega
_{\perp }$, which, after a simple analysis, leads to relation
\begin{equation}
2n=\left( (3/2)\pi ^{2}\rho _{\mathrm{3D}}a_{\mathrm{ho}}^{3}\right) ^{2/3}.
\label{n}
\end{equation}

One-dimensional function $\Phi (x,t)$ in ansatz (\ref{psiphi})\thinspace\
accounts for the dynamics in the $x$ direction, its normalization being
determined by Eq. (\ref{total}). The substitution of Eq. (\ref{psiphi}) in
Eq. (\ref{threeD1}) and subsequent averaging of the 3D equation in the
transverse plane yield the effective 1D equation
\begin{eqnarray}
i\hbar \Phi _{t} &=&-\frac{\hbar ^{2}}{2m_{\mathrm{eff}}}\Phi _{xx}+\left(
\frac{3}{2}\pi ^{2}\right) ^{2/3}\frac{3\hbar ^{2}}{{10}m}|\Phi |^{4/3}\Phi
\notag \\
&&+\frac{\pi a\hbar ^{2}}{m}|\Phi |^{2}\Phi -\epsilon \cos \left( \frac{4\pi
}{\lambda }x\right) \Phi ,  \label{1D}
\end{eqnarray}%
where the 1D potential with strength $-\epsilon $ corresponds to the OL with
period $\lambda /2$ (the negative sign in front of $\epsilon $ implies that
a local minimum of the potential is set at $x=0$, where the center of the
soliton will be placed).

By means of rescaling
\begin{equation}
\Phi \equiv \sqrt{\frac{2N}{n\lambda }}a_{\mathrm{ho}}^{-1}{\psi },~t=\frac{%
m_{\mathrm{eff}}\lambda ^{2}}{4\pi ^{2}\hbar }\tilde{t},~x\equiv \frac{%
\lambda }{2\pi }\tilde{x},  \label{scaling}
\end{equation}%
Eq. (\ref{1D}) is cast in the dimensionless form,
\begin{eqnarray}
i\psi _{t} &=&-\frac{1}{2}\psi _{xx}+G_{\mathrm{1D}}^{(7/3)}|\psi |^{4/3}\psi
\notag \\
&&+g_{\mathrm{1D}}|\psi |^{2}\psi -V_{0}\cos \left( 2x\right) \psi
\label{1Dequation}
\end{eqnarray}%
(tildes are dropped here), with the rescaled 1D wave function subject to
normalization
\begin{equation}
\int_{-\infty }^{+\infty }\left\vert \psi (x)\right\vert ^{2}dx=1.  \label{N}
\end{equation}%
The effective strengths of the Fermi nonlinearity, weak cubic interaction,
and OL potential are defined here as follows:
\begin{eqnarray}
G_{\mathrm{1D}}^{(7/3)} &\equiv &\frac{{3}}{{10}}\frac{m_{\mathrm{eff}}}{m}%
\left( \frac{{3}\lambda ^{2}N}{{8}\pi na_{\mathrm{ho}}^{2}}\right) ^{2/3},~~
\notag \\
V_{0} &\equiv &m_{\mathrm{eff}}\left( \frac{\lambda }{2\pi \hbar }\right)
^{2}\epsilon ,\quad ~g_{\mathrm{1D}}=\frac{m_{\mathrm{eff}}}{m}\frac{%
a\lambda N}{2\pi na_{\mathrm{ho}}^{2}}~.  \label{g}
\end{eqnarray}%
Actually, $V_{0}$ is the OL depth measured in units of the {recoil energy, $%
E_{R}=(2\pi \hbar )^{2}/(2m\lambda ^{2})$}. Below, it will be demonstrated
that $G_{\mathrm{1D}}^{(7/3)}$ takes values $\lesssim 10$ (see Fig. \ref%
{fig2}), while $g_{\mathrm{1D}}$ is confined to the range of $\sim 10^{-3}$,
hence the cubic term may be safely neglected in the first approximation.

\subsection{The two-dimensional equation}

Under the transverse confinement in direction $z$, which corresponds to the
``pancake" configuration, Eq. (\ref{threeD1}) can be reduced to a 2D form in
the plane of $\left( x,y\right) $. To this end, following \cite{Luca}, we
substitute $\Psi (x,y,z,t)=\Phi (x,y,t)\exp \left( -z^{2}/\left( 2na_{%
\mathrm{ho}}^{2}\right) \right) $, cf. Eq. (\ref{psiphi}). Averaging in $z$
and making use of the same rescalings as in Eq. (\ref{scaling}), except for
a different transformation of the wave function,
\begin{equation}
\Phi \equiv \frac{2\pi ^{3/4}\sqrt{N}}{n^{1/4}\lambda \sqrt{a_{\mathrm{ho}}}}%
{\psi }~,  \label{scaling2D}
\end{equation}%
we arrive at the following 2D equation (tildes are again dropped here):
\begin{eqnarray}
i\psi _{t} &=&-\frac{1}{2}\left( \partial _{x}^{2}+\partial _{y}^{2}\right)
\psi +G_{\mathrm{2D}}^{(7/3)}|\psi |^{4/3}\psi  \notag \\
&&+g_{\mathrm{2D}}|\psi |^{2}\psi -V_{0}\left[ \cos \left( 2x\right) +\cos
\left( 2y\right) \right] \psi ,  \label{2Dequation}
\end{eqnarray}%
where we have included the 2D lattice potential and adopted the following
definitions:
\begin{eqnarray}
G_{\mathrm{2D}}^{(7/3)} &\equiv &\frac{1}{{\ \sqrt{5}}}\left( \frac{{3^{7/2}}%
\pi }{{64}n}\right) ^{1/3}\frac{m_{\mathrm{eff}}}{m}\left( \frac{\lambda }{%
a_{\mathrm{ho}}}N\right) ^{2/3},~  \notag \\
g_{\mathrm{2D}} &\equiv &2\sqrt{2\pi /n}\frac{m_{\mathrm{eff}}}{m}\frac{a}{%
a_{\mathrm{ho}}}N,  \label{g2D}
\end{eqnarray}%
\begin{equation}
\int \int |\psi (x,y)|^{2}dxdy=1.  \label{N2D}
\end{equation}%
Finally, the above-mentioned self-consistency condition, $n\hbar \omega
_{\perp }=\varepsilon _{F}$, leads, in the present case, to relation%
\begin{equation}
2n=\left( \frac{3}{\sqrt{2}}\pi ^{2}\rho _{\mathrm{3D}}a_{\mathrm{ho}%
}^{3}\right) ^{2/3},  \label{n2D}
\end{equation}%
cf. its counterpart (\ref{n}) for the cigar-shaped configuration.

\subsection{Other equations}

The derivation presented above should be modified in the case of extremely
tight transverse confinement, with $\varepsilon _{F}\ll \hbar \omega _{\perp
}$. If the very tight trap is cigar-shaped, the derivation must start from
the 1D Fermi distribution, with the respective atomic density related to $%
\varepsilon _{F}$ as follows: $\rho _{\mathrm{1D}}={2(2\pi )}^{-1}$ $%
\int_{-k_{F}}^{+k_{F}}dk\equiv \left( 2/\pi \hbar \right) \sqrt{%
2m\varepsilon _{F}}$, hence
\begin{equation}
\varepsilon _{F}=\pi ^{2}\hbar ^{2}\rho _{1D}^{2}/(8m).  \label{Fermi1D}
\end{equation}%
It is known that the energy density of the 1D Fermi superfluid is \cite%
{Yang1D} $\mathcal{E}_{\mathrm{1D}}=(1/3)\rho _{\mathrm{1D}}\varepsilon _{F}$%
. 
Using expression (\ref{Fermi1D}), one can derive the energy density of the
1D superfluid \cite{Tokatly}:
\begin{equation}
\mathcal{E}_{\mathrm{1D}}=\left( \pi ^{2}\hbar ^{2}/24m\right) \rho _{%
\mathrm{1D}}^{3}.  \label{1G}
\end{equation}%
Similarly, for the 2D superfluid subjected to an extremely tight confinement
in direction $z$, the energy density is $\mathcal{E}_{\mathrm{2D}}=(1/2)\rho
_{\mathrm{2D}}\varepsilon _{F}$, the 2D atomic density being $\rho _{\mathrm{%
2D}}={2(2\pi )^{-2}}\int_{0}^{k_{F}}2\pi kdk\equiv \left( m/\pi \hbar
^{2}\right) \varepsilon _{F}$, i.e., $\varepsilon _{F}=\pi \hbar ^{2}\rho _{%
\mathrm{2D}}/m$. Thus, the energy density of the 2D superfluid can be
written in terms of the 2D density \cite{luca2},
\begin{equation}
\mathcal{E}_{\mathrm{2D}}=\left( \pi \hbar ^{2}/2m\right) \rho _{\mathrm{2D}%
}^{2}~.  \label{1J}
\end{equation}

Repeating the analysis which produced Eq. (\ref{threeD1}), we arrive at the
following 2D and 1D equations corresponding to the limit case of the very
tight transverse trap:
\begin{eqnarray}
i\hbar \psi _{t} &=&\frac{\hbar ^{2}}{2}\left\{
\begin{array}{c}
-\left( 1/m_{\mathrm{eff}}\right) \left( \partial _{x}^{2}+\partial
_{y}^{2}\right) +\left( 2\pi /m\right) |\psi |^{2} \\
-\left( 1/m_{\mathrm{eff}}\right) \partial _{x}^{2}+(\pi ^{2}/4m)|\psi |^{4}%
\end{array}%
\right\} \psi  \notag \\
&&+V(\mathbf{r})\psi ~.  \label{one/twoD}
\end{eqnarray}%
The equation with the cubic nonlinearity, in the first line, is formally
equivalent to the ordinary GPE in two dimensions, while the 1D equation in
the second line coincides with the above-mentioned quintic equation for the
1D gas of hard-core bosons \cite{further}.

Assuming that $V$ represents the OL potential in the 1D variant of Eq. (\ref%
{one/twoD}), the respective families of GS solutions \cite{we} are
mathematically tantamount to the 1D bosonic GSs generated by the
above-mentioned NLS equation with the quintic self-repulsive term \cite%
{Fatkhulla}. However, unlike Eq. (\ref{1Dequation}), the physical relevance
of the 1D variant of Eq. (\ref{one/twoD}) for the prediction of GSs in the
BCS superfluid is impugnable, because subsequent analysis demonstrate that
such solitons, on the contrary to those produced by Eq. (\ref{1Dequation}),
contain few ($\ll 100$) atoms \cite{we}, which makes the very concept of the
superfluid doubtful in such a situation. Similarly, it is possible to
construct a family of GSs in the 2D version of Eq. (\ref{one/twoD}),
assuming that $V(\mathbf{r})$ is the 2D OL potential. Mathematically, they
will be equivalent to the GS solutions of the ordinary 2D GPE \cite%
{GSprediction}. However, in this case too, the number of atoms in such 2D
solitons, if they are realized in terms of the BCS superfluid, turns out to
be small (on the contrary to the 2D solitons predicted by Eq. (\ref%
{2Dequation})). Therefore, we will focus on the physically relevant models
based on Eqs. (\ref{1Dequation}) and (\ref{2Dequation}). The results will be
compared to those for the bosonic GSs generated by one- and two-dimensional
GPEs, (\ref{bosonic}) and (\ref{2Dbosonic}), see below.

\section{One-dimensional solitons}

\subsection{Variational approximation}

Stationary solutions to Eq. (\ref{1Dequation}) are looked for in the usual
form, $\psi (x,t)=e^{-i\mu t}\phi (x)$, where $\mu $ is the chemical
potential, and real function $\phi (x)$ obeys equation%
\begin{equation}
\mu \phi +(1/2)\phi ^{\prime \prime }-G_{\mathrm{1D}}^{(7/3)}\phi
^{7/3}+V_{0}\cos \left( 2x\right) \phi =0,  \label{phi1D}
\end{equation}%
with $\phi ^{\prime \prime }\equiv d^{2}\phi /dx^{2}$, in which the small
cubic term is dropped (its effect will be considered below). This equation,
together with normalization condition (\ref{N}), can be derived from the
following Lagrangian,
\begin{eqnarray}
L &=&\int_{-\infty }^{+\infty }\left[ \mu \phi ^{2}-\frac{1}{2}\left( \phi
^{\prime }\right) ^{2}-\frac{3}{5}G_{\mathrm{1D}}^{(7/3)}\phi ^{10/3}\right.
\notag \\
&&\left. +V_{0}\cos (2x)\phi ^{2}\right] dx-\mu ,  \label{L}
\end{eqnarray}%
by demanding $\delta L/\delta \phi =\partial L/\partial \mu =0$. To apply
the VA, we use the Gaussian ansatz \cite{VA},
\begin{equation}
\phi (x)=\pi ^{-1/4}\sqrt{\frac{M}{W}}\exp \left( -\frac{x^{2}}{2W^{2}}%
\right) ,  \label{Gauss}
\end{equation}%
where variational parameters are the soliton's norm and width, $M$ and $W$
(in addition to $\mu $). The use of the Gaussian is justified by the
numerical results showing that, except for a narrow vicinity of bandgap
edges, the fundamental GSs are strongly localized solutions, see below. We
also note that ansatz (\ref{Gauss}) implies that the center of the soliton
is placed at a local minimum of the OL potential. A straightforward
generalization of the VA, which treats the central coordinate as another
degree of freedom of the GS, demonstrates that solitons may also be found
with the center located at a local potential maximum, but they are always
unstable against the shift from that position.

The substitution of ansatz (\ref{Gauss}) in Lagrangian (\ref{L}) yields
\begin{eqnarray}
L &=&\mu \left( M-1\right) -\frac{M}{4W^{2}}+V_{0}Me^{-W^{2}}  \notag \\
&&-\left( \frac{3}{5}\right) ^{3/2}\frac{G_{\mathrm{1D}}^{(7/3)}M^{5/3}}{\pi
^{7/6}W^{2/3}}.  \label{LGauss}
\end{eqnarray}

The first variational equation following from Eq. (\ref{LGauss}), $\partial
L/\partial \mu =0$, yields (as it should) $M=1$, therefore, $M=1$ is
substituted in other variational equations below, except for equation $%
\partial L/\partial M=0$, where $M=1$ is substituted after the
differentiation. The remaining variational equations are $\partial
L/\partial W=0$, which predicts a relation between the soliton's width and
effective nonlinearity strength, $G_{\mathrm{1D}}^{(7/3)}$,
\begin{equation}
1+\frac{4\sqrt{3}G_{\mathrm{1D}}^{(7/3)}}{5^{3/2}\pi ^{1/3}}%
W^{4/3}=4V_{0}W^{4}e^{-W^{2}},  \label{WGauss}
\end{equation}%
and $\partial L/\partial M=0$, which yields $\mu $ as a function of $W$ and $%
G_{\mathrm{1D}}^{(7/3)}$:
\begin{equation}
\mu =\frac{1}{4W^{2}}+\frac{\sqrt{3}G_{\mathrm{1D}}^{(7/3)}}{\pi ^{1/3}\sqrt{%
5}W^{2/3}}-V_{0}e^{-W^{2}}.  \label{muGauss}
\end{equation}

The first noteworthy consequence of the above equations is that, for $V_{0}$
not too small, they predict a certain value of $\mu $ at $G_{\mathrm{1D}%
}^{(7/3)}\rightarrow 0$, which corresponds to the stationary \emph{linear}
Schr\"{o}dinger equation. Indeed, setting $G_{\mathrm{1D}}^{(7/3)}=0$ in Eq.
(\ref{WGauss}) yields%
\begin{equation}
4V_{0}=W^{-4}\exp \left( W^{2}\right) .  \label{artifact}
\end{equation}%
This expression attains a minimum,
\begin{equation}
\left( V_{0}\right) _{\min }=e^{2}/16\approx \allowbreak 0.462,  \label{min}
\end{equation}%
at $W^{2}=2$. At $V_{0}>\left( V_{0}\right) _{\min }$, Eq. (\ref{artifact})
yields two solutions for $W^{2}$.

\begin{figure}[tbp]
{\includegraphics[width=1.0\linewidth,clip]{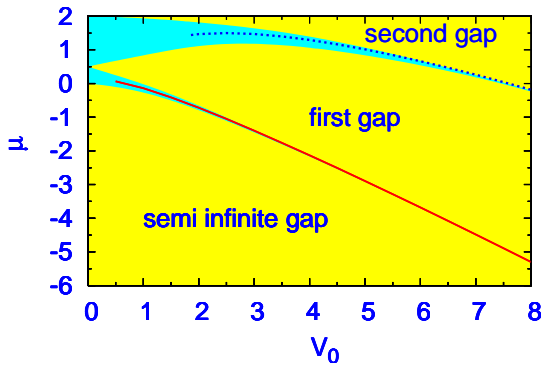}}
\caption{(Color online) The bandgap structure of the linearized version of
Eq. (\protect\ref{phi1D}). Narrow regions between the gaps represent Bloch
bands. The solid and dotted curves are borders between the semi-infinite and
first finite gaps, and between the first and second finite gaps, as
predicted by the variational approximation (see text).}
\label{fig1}
\end{figure}

A commonly known fact is that the linear Schr\"{o}dinger equation with the
periodic potential cannot give rise to localized states. Nevertheless, the
results presented in Fig. \ref{fig1}, which displays the well-known band
structure for the linear version of Eq. (\ref{phi1D}), together with the $%
\mu \left( V_{0}\right) $ curve (the continuous red line), as obtained from
Eqs. (\ref{artifact}) and (\ref{muGauss}) with $G_{\mathrm{1D}}^{(7/3)}=0$,
clearly indicate that, if $V_{0}$ exceeds minimum value (\ref{min}), the
variational solution makes sense in the linear limit: it accurately predicts
the location of the left/lower edge of the first finite bandgap, as the
starting point of the GS family inside this bandgap, see Fig. \ref{fig2}
below. In other words, Fig. \ref{fig1} shows that the VA makes it possible
to predict the location of the narrow Bloch band which separates the
semi-infinite gap and first finite bandgap (a similar result was obtained by
means of VA in \cite{GMM}). In particular, in the case of $V_{0}=5$, for
which most results are presented below, Eqs. (\ref{artifact}) and (\ref%
{muGauss}) with $G_{\mathrm{1D}}^{(7/3)}=0$ predict the left edge of the
first finite bandgap at
\begin{equation}
\left( \mu _{1}\right) _{\mathrm{left}}^{\mathrm{(var)}}(V_{0}=5)\approx
-2.894.  \label{var}
\end{equation}%
This value almost exactly coincides with its counterpart found from the
numerical solution of  Mathieu equation (i.e., the linearization of Eq. 
(\ref%
{phi1D}), $\left( \mu _{1}\right) _{\mathrm{left}}^{\mathrm{(num)}%
}(V_{0}=5)\approx -2.893$.

The above-mentioned variational curve shown in Fig. \ref{fig1} corresponds
to a smaller root of Eq. (\ref{artifact}). Through the other (larger) root
of the same equation, the VA is actually trying to predict another narrow
Bloch band, which separates the first and second finite bandgaps in Fig. \ref%
{fig1}. This root produces a large error in predicting the border between
the first and second gaps, as the underlying even ansatz, adopted in Eq. (%
\ref{Gauss}), is inadequate in that case. However, the border can be
accurately predicted by a modified version of the VA, based on a properly
modified (odd) ansatz, see Eq. (\ref{Gauss1}) in Appendix and the dotted
blue curve in Fig. \ref{fig1}.

In addition to the VA based on Gaussian ansatz (\ref{Gauss}), we have also
elaborated its counterpart based on the hyperbolic secant, $\phi (x)=\sqrt{%
M/2W}\mathrm{sech}\left( x/W\right) $. Eventual results produced by this
ansatz, which we do not display here, are quite close to those generated by
the Gaussian, although the accuracy is slightly lower; in particular, it
predicts $\left( \mu _{1}\right) _{\mathrm{left}}^{\mathrm{(var)}%
}(V_{0}=5)-2.854$, cf. value (\ref{var}) produced by the Gaussian-based VA.

\subsection{Numerical results: fundamental gap solitons}

Numerical solutions where obtained by integration of time-dependent Eq. (\ref%
{1Dequation}) (and time-dependent counterparts of bosonic Eqs. (\ref{bosonic}%
), see below), using the Crank-Nicholson discretization scheme, until the
solution would converge to a time-independent real form. The equations were
discretized using time step $0.0005$ and space step $0.025$, in domain $%
-20<x<+20$. This method of obtaining the stationary solutions guarantees
that they are stable. The results presented here were obtained dropping the
small cubic term in Eq. (\ref{1Dequation}); effects induced by this
additional term are considered below separately.

\begin{figure}[tbp]
\begin{center}
{\includegraphics[width=1.0\linewidth]{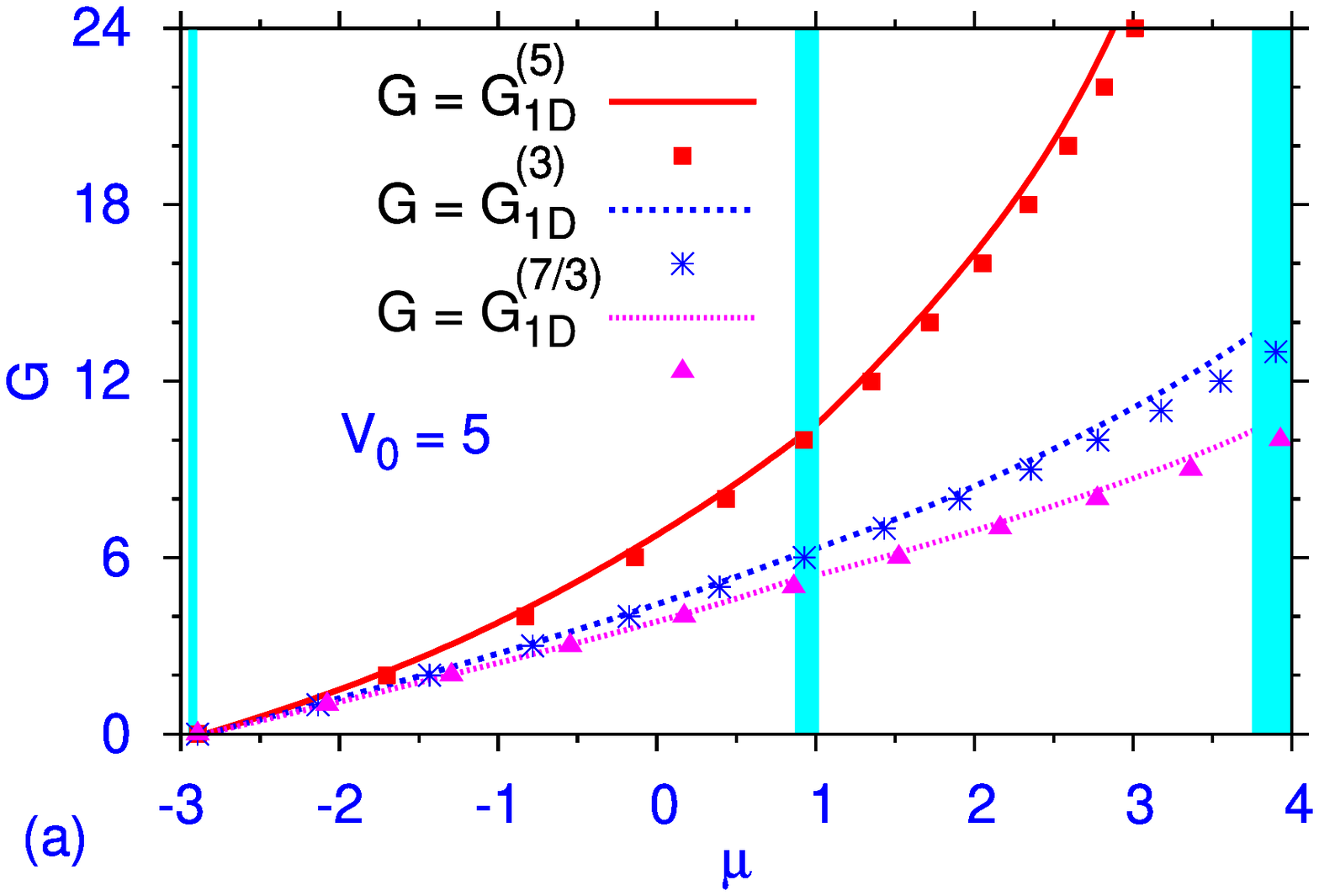}} {%
\includegraphics[width=1.0\linewidth]{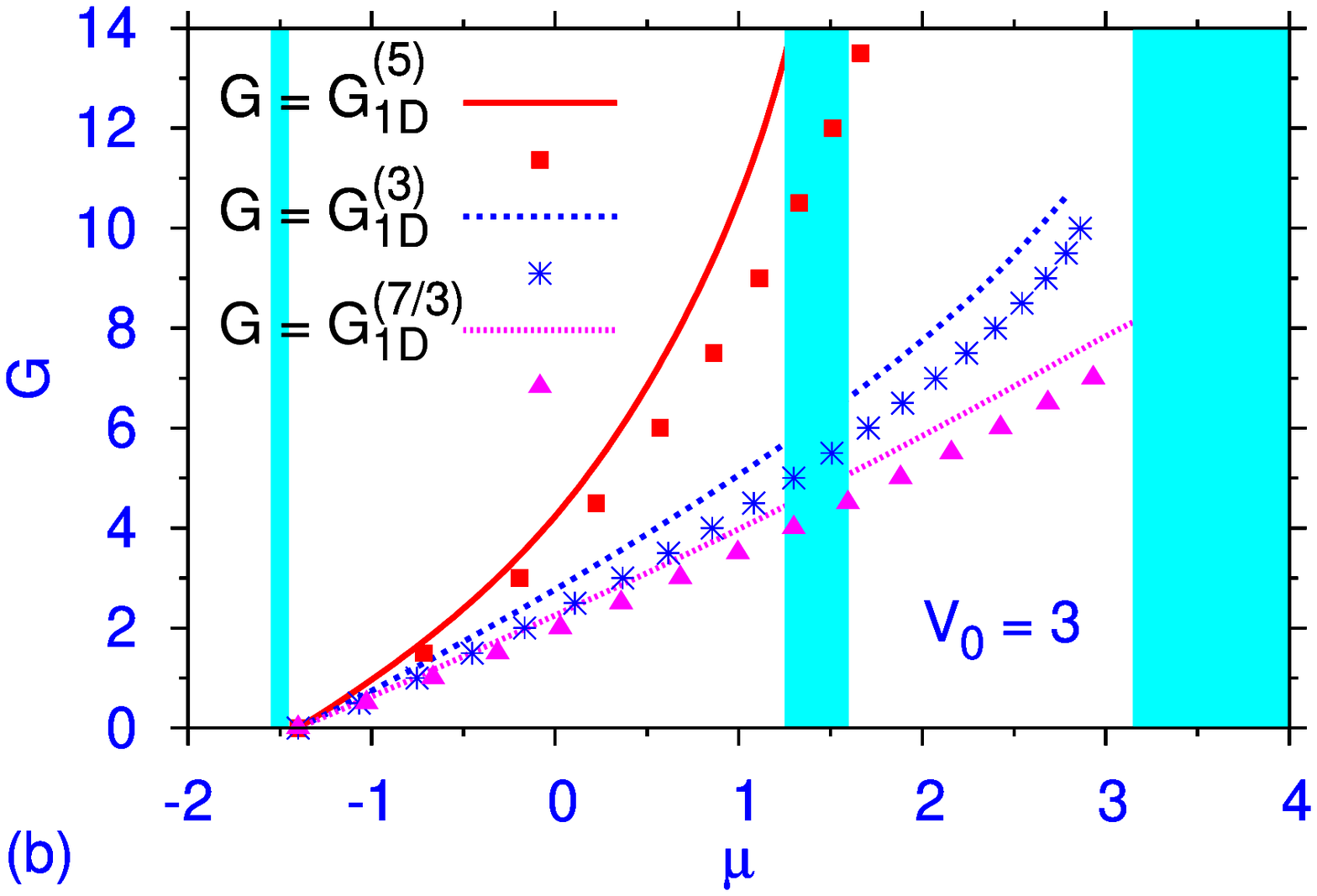}}
\end{center}
\caption{(Color online) The nonlinearity coefficient, $G_{\mathrm{1D}%
}^{(7/3)}$, in Eq. (\protect\ref{1Dequation}) for the BCS superfluid, versus
chemical potential $\protect\mu $, for the family of fundamental 1D gap
solitons found in the first two finite bandgaps of periodic potential $%
-V_{0}\cos (2x)$, with $V_{0}=5$ (a) and $V_{0}=3$ (b). For comparison, also
shown are dependences between the respective nonlinear coefficient, $G_{%
\mathrm{1D}}^{(3)}$ or $G_{\mathrm{1D}}^{(5)}$, and $\protect\mu $ for
families of fundamental gap solitons in two 1D bosonic equations (\protect
\ref{bosonic}). Here and in similar figures displayed below, shaded vertical
areas represent Bloch bands which separate the gaps. Numerically found
solution families are depicted by continuous curves, whereas the predictions
produced by the variational approximation are shown by chains of symbols.}
\label{fig2}
\end{figure}

The family of fundamental GSs predicted by the VA is characterized by
dependence $G_{\mathrm{1D}}^{(7/3)}(\mu )$, which was obtained from a
numerical solution of Eqs. (\ref{WGauss}) and (\ref{muGauss}) (it can be
easily translated into a dependence between $\mu $ and the number of atoms,
using Eqs. (\ref{g}) and (\ref{n})). This dependence is displayed for two
different values of the OL strength, $V_{0}$, in Fig. \ref{fig2}. For the
sake of comparison, the figure includes similar dependences for families of
1D bosonic GSs, as obtained by means of the VA, based on the same ansatz (%
\ref{Gauss}), from the following bosonic equations with the cubic and
quintic self-repulsive terms,
\begin{equation}
\mu \phi +\frac{1}{2}\phi ^{\prime \prime }-\left\{
\begin{array}{c}
G_{\mathrm{1D}}^{(3)}\phi ^{3} \\
G_{\mathrm{1D}}^{(5)}\phi ^{5}%
\end{array}%
\right\} +V_{0}\cos (2x)\phi =0.~  \label{bosonic}
\end{equation}%
Here, wave function $\phi (x)$ is subject to normalization condition (\ref{N}%
), and $G_{\mathrm{1D}}^{(3)}=\left( \lambda a_{s}/\pi a_{\mathrm{ho}%
}^{2}\right) N$ \cite{GSprediction,Luca}, $G_{\mathrm{1D}}^{(5)}=\left( \pi
^{2}/2\right) $ $N^{2}$ \cite{further}, with $N$ the number of boson atoms, $%
a_{s}$ the scattering length characterizing the repulsive interactions
between them, and $a_{\mathrm{ho}}$ the same transverse-trapping size as
above.

\begin{figure}[tbp]
{\includegraphics[width=1.0\linewidth,clip]{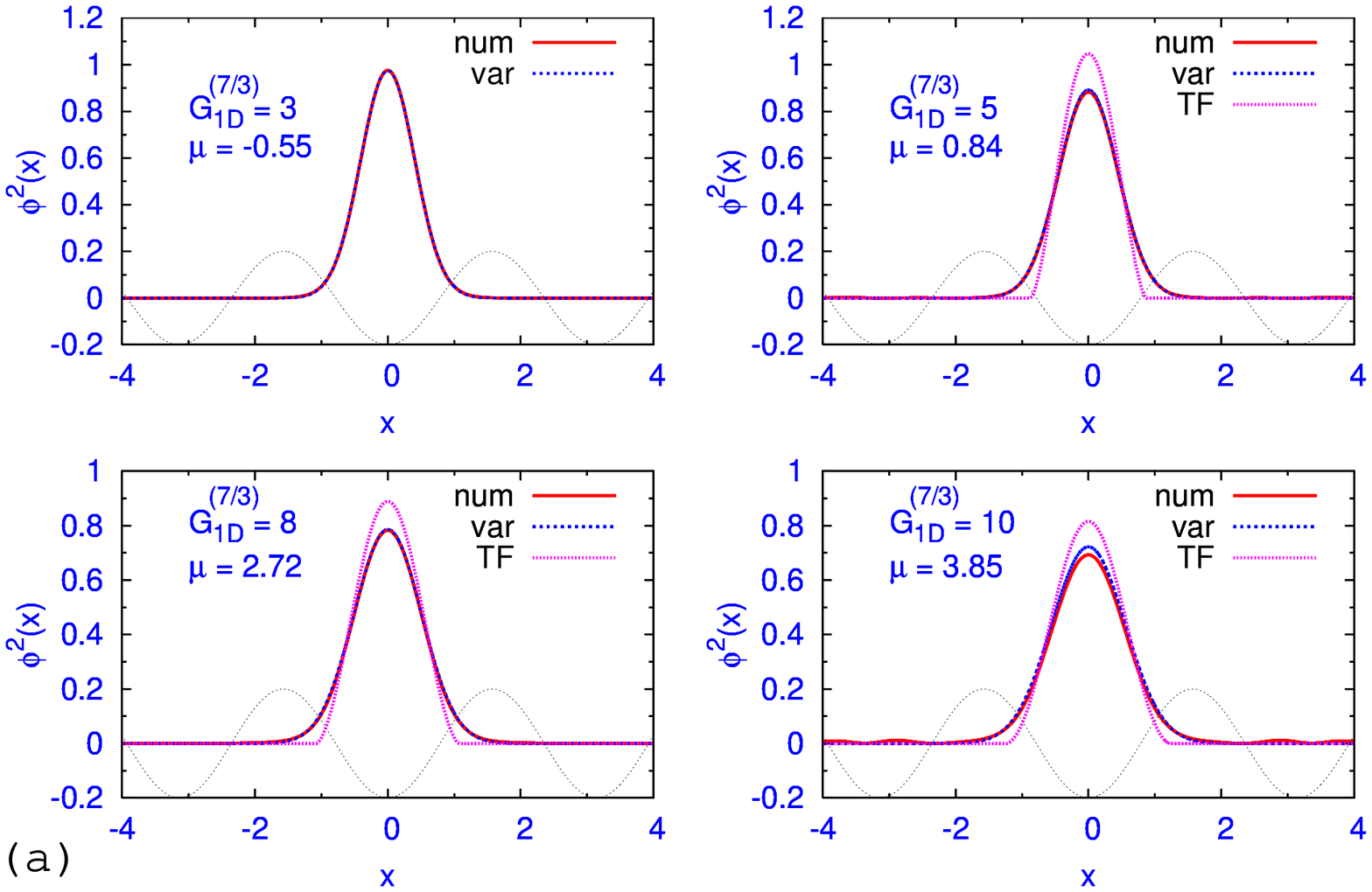}} {%
\includegraphics[width=1.0\linewidth,clip]{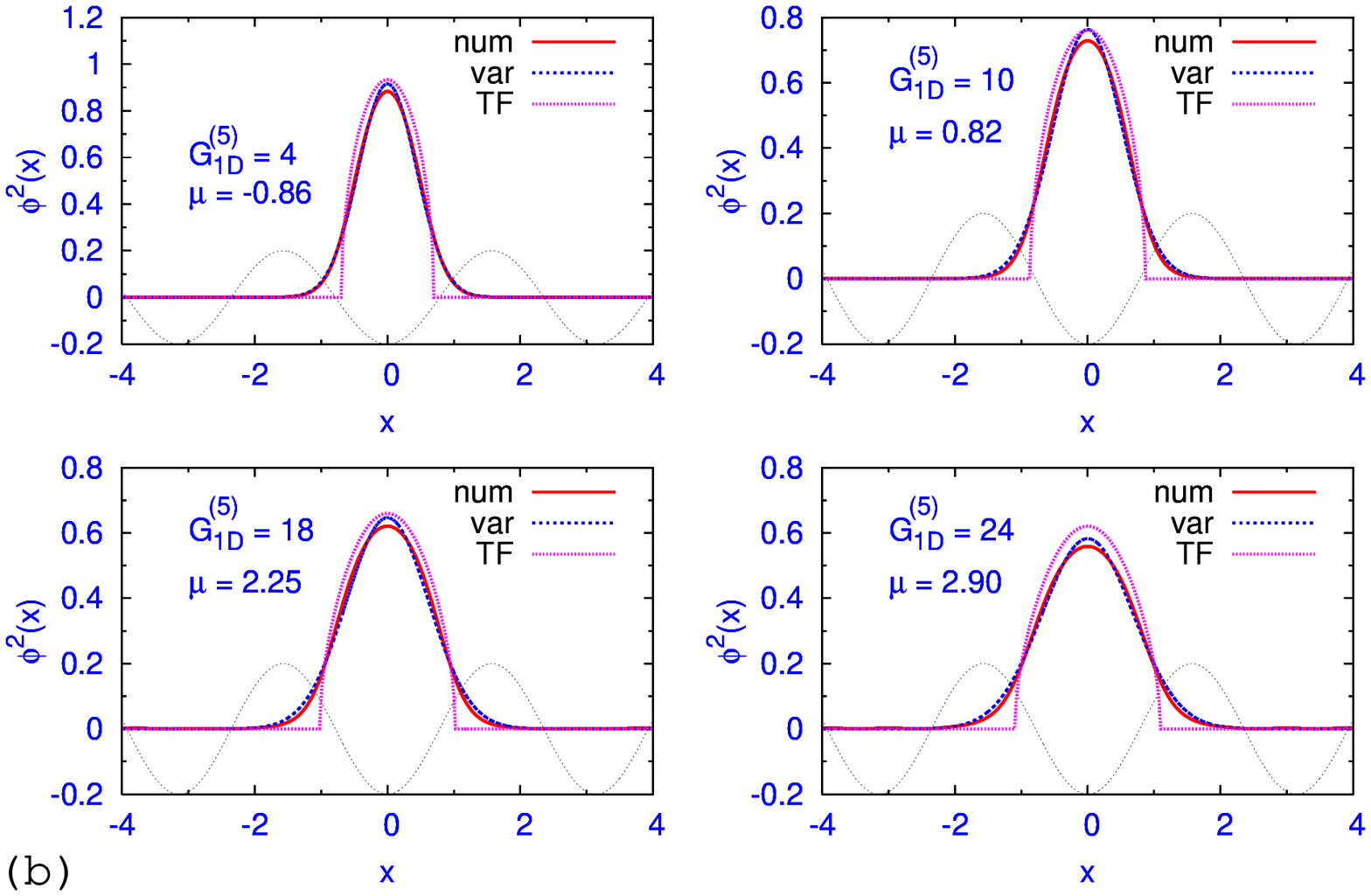}}
\caption{(Color online) One-dimensional GS wave forms (normalized to unity)
in the first and second bandgaps as obtained (a) from Eq. (\protect\ref%
{phi1D}), without the small cubic term, and (b) from bosonic Eq. (\protect
\ref{bosonic}) with the quintic nonlinearity, for $V_{0}=5$. Here and below,
labels \textquotedblleft num", \textquotedblleft var", and \textquotedblleft
TF" represent numerical, variational, and Thomas-Fermi results, respectively
and the thin sinusoidal line depicts the underlying periodic potential, $%
-V_{0}\cos (2x)$.}
\label{fig3}
\end{figure}

Comparison of typical shapes of the numerically found GSs with the
respective VA profiles predicted by the Gaussian ansatz is displayed in Fig. %
\ref{fig3} (a set of GS shapes obtained from Eq. (\ref{bosonic}) with the
quintic nonlinearity, which is also displayed in this figure, makes it
possible to compare the GSs in the BCS superfluid with their bosonic
counterparts). The figure includes examples of the GSs belonging to both the
first and second bandgaps, which can be easily identified by values of the
chemical potential indicated in the panels. In addition to the numerical and
variational profiles, in Fig. \ref{fig3} we also plot their simplest
counterparts predicted by the Thomas-Fermi (TF) approximation, which were
obtained, as usual \cite{books}, by dropping the second-derivative term in
Eqs. (\ref{phi1D}) or (\ref{bosonic}) (the TF profiles are not shown for
very weak nonlinearity, where this approximation is irrelevant)

Figure \ref{fig3} demonstrates that the fundamental GSs, unless taken near
edges of the bandgaps, are indeed compact objects with a quasi-Gaussian
shape, trapped in a single cell of the OL potential. This shape radically
changes as one comes very close to a bandgap's edge, or pushes the solutions
to higher values of $\mu $ (in particular, the curves representing the GS
family in Eq. (\ref{bosonic}) with the cubic nonlinearity are aborted in
Fig. \ref{fig2} (b) at a spot where the soliton undergoes a transition to a
complex shape with undulating tails, making the further use of the VA
irrelevant).

The variational solutions displayed in Fig. \ref{fig2}(a) emerge, at $G=0$,
at the value of $\mu $ given by Eq. (\ref{var}) (recall that it almost
exactly coincides with the actual left edge of the first bandgap; the same
is true for Fig. \ref{fig2}(b). However, an essential defect of the
variational solutions for these families is that they ignore the Bloch band
separating the first and second bandgaps, going across it (inside the bands
the VA predicts the so-called quasi-solitons, i.e., nearly localized
solutions, with the smallest amplitude of nonvanishing tails attached to the
central \textquotedblleft body" \cite{Dave}). Another noteworthy inference
is that, as seen from Figs. \ref{fig2} and \ref{fig3}, the accuracy provided
by the VA for the description of the GS family generated by Eq. (\ref{phi1D}%
) for the BCS superfluid is \emph{better} than for the bosonic GS families.

\subsection{Physical estimates for the fundamental gap solitons}

Undoing rescalings (\ref{scaling}) and (\ref{g}), we conclude that
conditions (\ref{de Broglie}) and (\ref{semiclassical}), which underlie the
derivation of the effective 1D Eq. (\ref{1Dequation}), are definitely
satisfied for $G_{\mathrm{1D}}^{(7/3)}>3$, i.e., according to Fig. \ref{fig2}%
, starting from the middle of the first finite bandgap. To assess the
feasibility of the creation of the GSs in the BCS superfluid trapped in the
OL, it is necessary to estimate the expected number of atoms, $N$, in the
predicted quasi-1D soliton. Getting back to the physical units, we arrive at
the following estimates for $N$ and the corresponding value of $n$ (recall
it is the largest quantum number of the filled states in the transverse
trapping potential, which appears in Eqs. (\ref{psiphi}) and (\ref{n})):%
\begin{eqnarray}
N_{\mathrm{1Dsoliton}} &\simeq &10^{3}\cdot \left( \frac{m}{m_{\mathrm{eff}}}%
\right) ^{3/2}\left( G_{\mathrm{1D}}^{(7/3)}\right) ^{5/2}\left( a_{\mathrm{%
ho}}/\lambda \right) ^{4},  \notag \\
n^{5} &\simeq &10\left( N_{\mathrm{1Dsoliton}}a_{\mathrm{ho}}/\lambda
\right) ^{2}.  \label{Nn}
\end{eqnarray}%
Figure \ref{fig2} demonstrates that the largest achievable value of the
effective nonlinearity strength is $\left( G_{\mathrm{1D}}^{(7/3)}\right)
_{\max }\sim 10$. Then, assuming $a_{\mathrm{ho}}/\lambda \sim 1$ (for $^{6}$%
Li atoms, this implies the use of trapping frequency $\ \omega _{\perp
}\simeq 2\pi \times 1$~KHz, if the OL is induced by light with wavelength $%
\lambda \simeq 1$ $\mu $m), relations (\ref{Nn}) predict $N_{\mathrm{%
1Dsoliton}}$ in the range of $10^{4}$, with $n\sim 100$. For a larger
wavelength, $\lambda \simeq 3$ $\mu $m, one concludes that $N_{\mathrm{%
1Dsoliton}}$ takes values in the range of $10^{3}$, with $n$ $\sim
\allowbreak 20$.

\subsection{Bound states of gap solitons}

\begin{figure}[tbp]
\begin{center}
{\includegraphics[width=1.0\linewidth,clip]{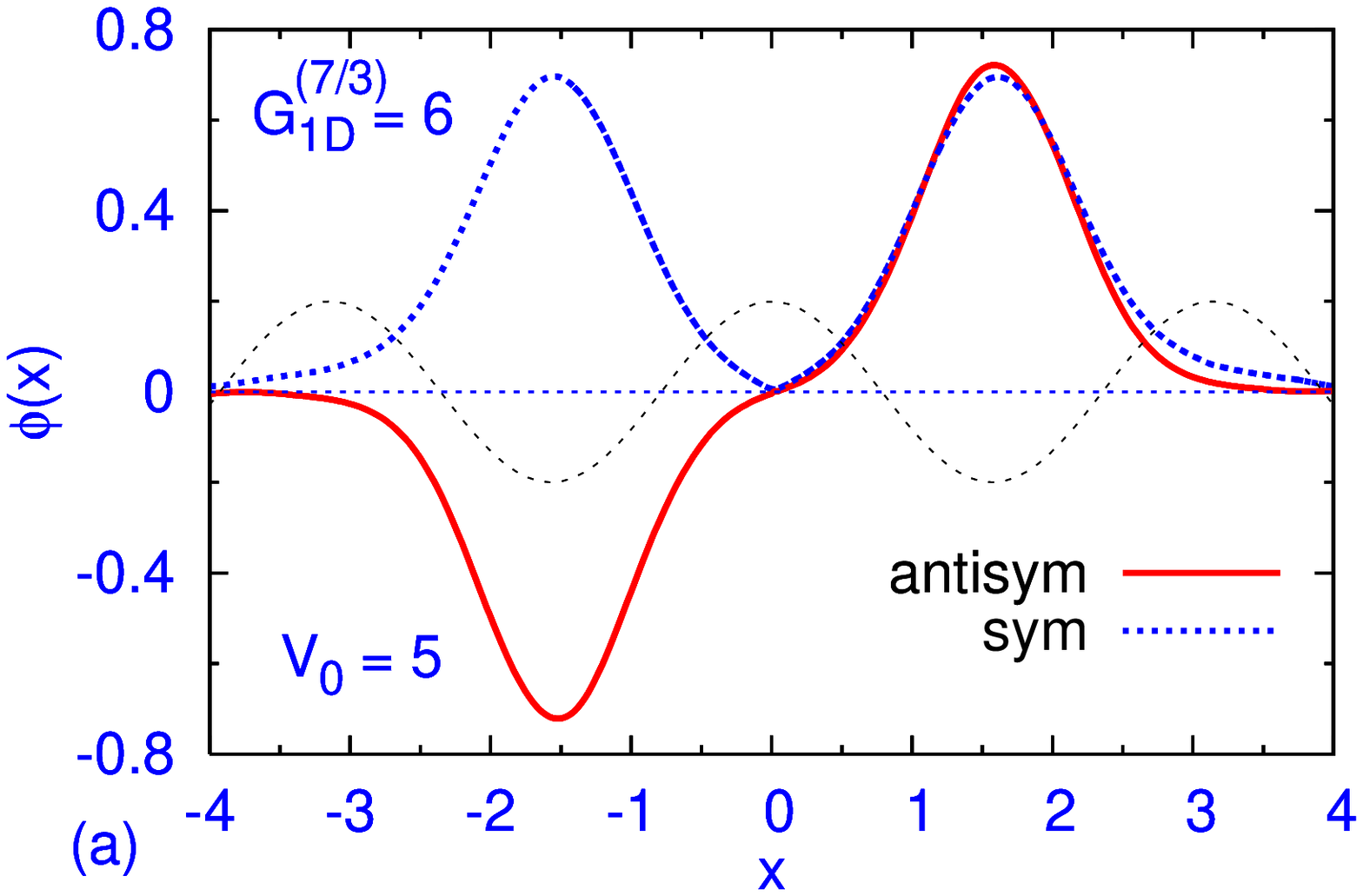}} {\ %
\includegraphics[width=1.0\linewidth,clip]{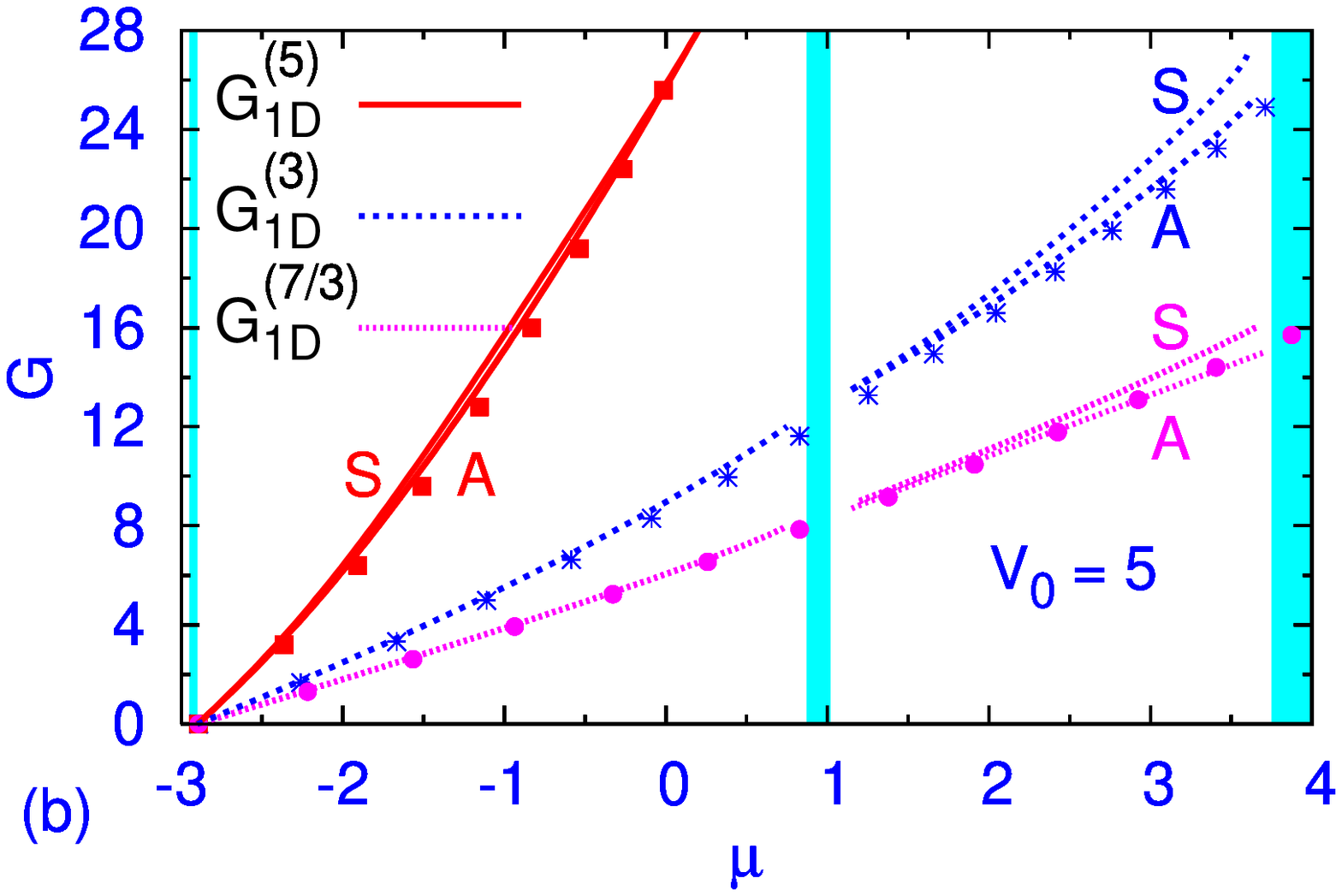}}
\end{center}
\caption{(Color online) (a) Examples of symmetric and antisymmetric
(\textquotedblleft twisted") stable bound states of two 1D fundamental GSs,
obtained from Eq. (\protect\ref{1Dequation}). (b) Numerically constructed
(continuous lines) plots of the nonlinearity versus the chemical potential
for symmetric (S) and antisymmetric (A) bound states, for the BCS superfluid
and two bosonic models. The analytical results (chains of symbols) are
generated using Fig. \protect\ref{fig2}, with $G_{\mathrm{1D}}$ rescaled as
described in the text. }
\label{fig6}
\end{figure}

We have also constructed symmetric and antisymmetric (\textquotedblleft
twisted") bound states of the fundamental GSs (unlike the SF solutions,
considered in Appendix, they are stable solutions to Eq. (\ref{1Dequation}%
)). These states were formed by placing in-phase or out-of-phase pairs of
identical GSs (ones with equal or opposite signs, respectively) in adjacent
cells of the OL potential and allowing them to achieve a stable
configuration. Typical examples are shown in Fig. \ref{fig6} (a). In the
experiment, the relative phase of adjacent atomic clusters may be controlled
by laser beams.

The $G_{\mathrm{1D}}(\mu )$ curves for the symmetric and antisymmetric bound
states are shown in Fig. \ref{fig6}(b) (again, they are presented together
with similar dependences for bound GS pairs generated by two bosonic
equations (\ref{bosonic})). As might be expected, these curves can be
obtained, in an almost exact form, from those for the fundamental solitons
(see Fig. \ref{fig2}), by means of rescalings: $G_{\mathrm{1D}%
}^{(7/3)}\rightarrow 2^{2/3}G_{\mathrm{1D}}^{(7/3)}$, $G_{\mathrm{1D}%
}^{(3)}\rightarrow 2G_{\mathrm{1D}}^{(3)}$, $G_{\mathrm{1D}%
}^{(5)}\rightarrow 4G_{\mathrm{1D}}^{(5)}$, respectively. Indeed, if the
equations are written in the notation with a fixed nonlinearity coefficient
and variable norm, the rescalings simply imply that the norm of the bound
state is twice that of the fundamental soliton. The fact that the so defined
norm of the symmetric bound states slightly exceeds the norm of their
antisymmetric counterparts is natural too, as the vanishing of the density
at the center of the antisymmetric state makes its total norm slightly
smaller.

\subsection{Dragging gap solitons by a moving lattice}

\begin{figure}[tbp]
\begin{center}
\includegraphics[width=.48\linewidth]{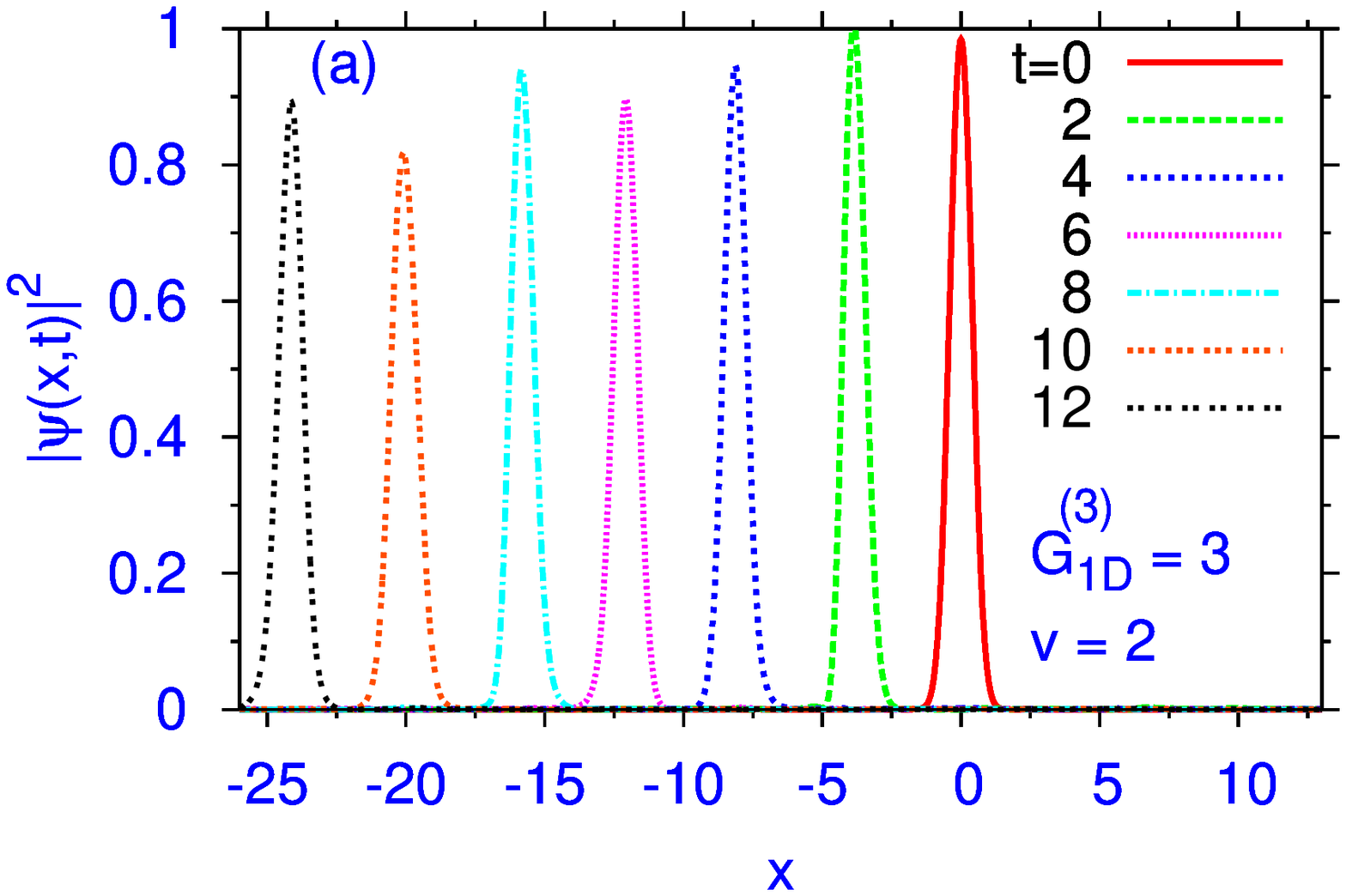} \includegraphics[width=.48%
\linewidth]{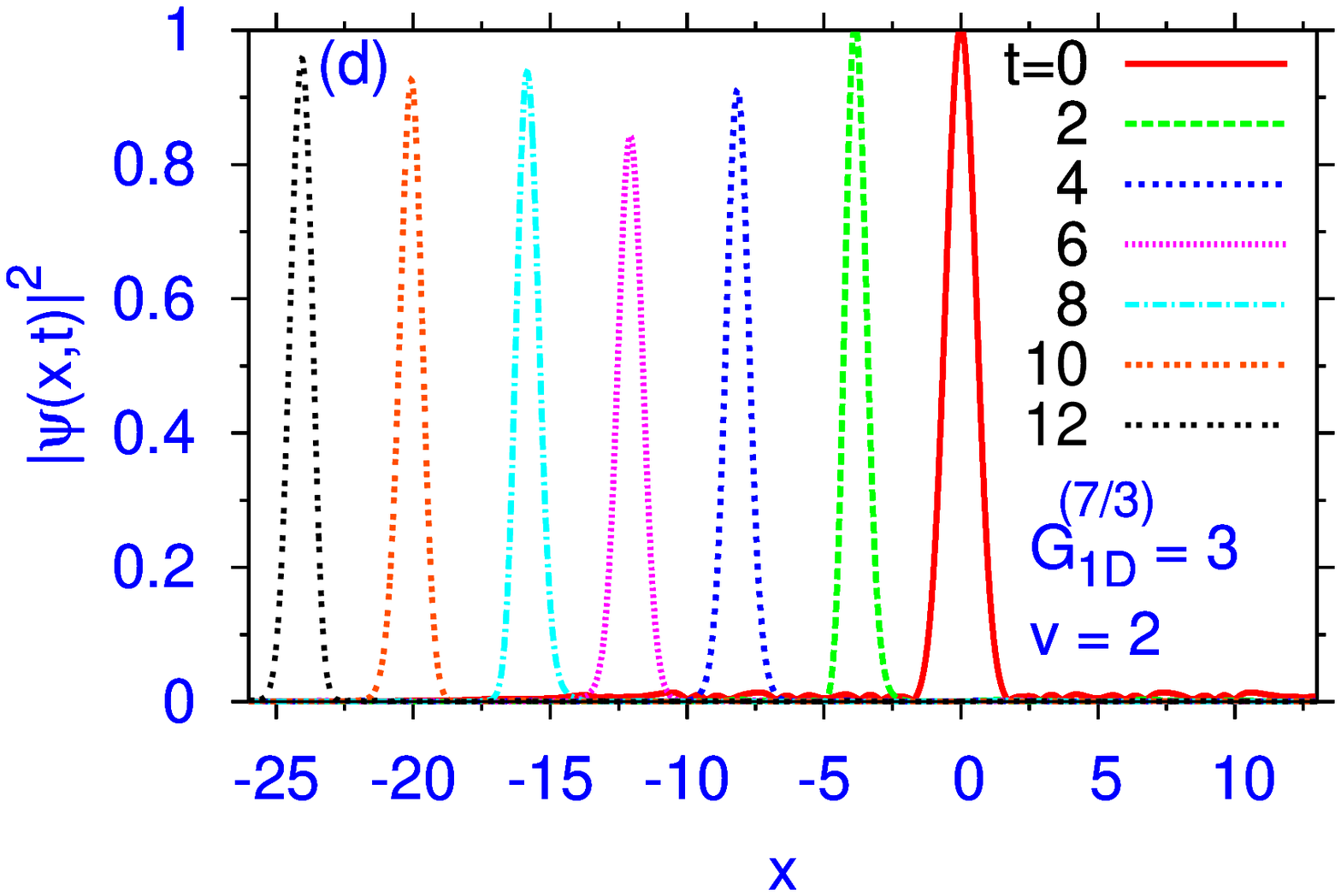} \includegraphics[width=.48\linewidth]{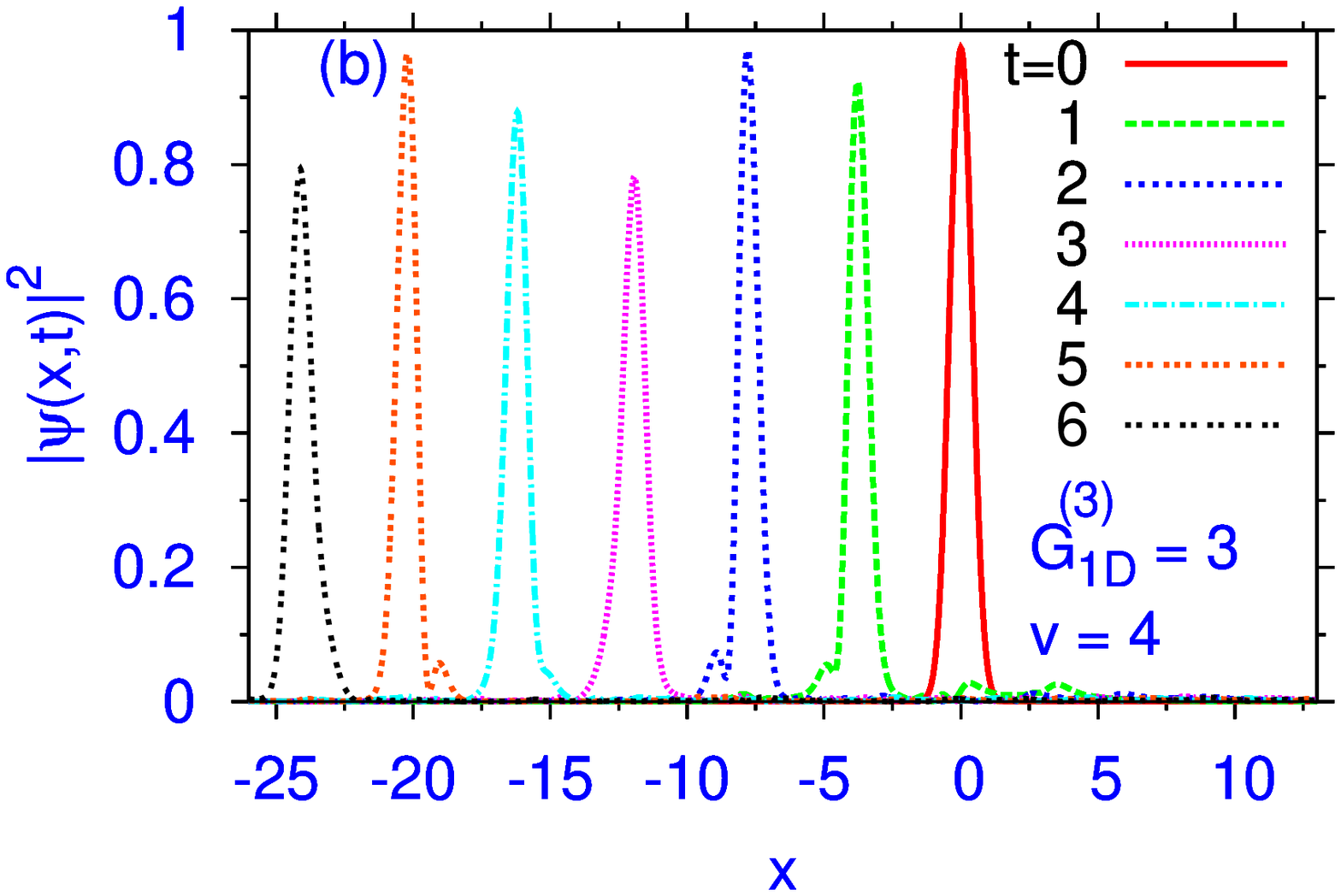} %
\includegraphics[width=.48\linewidth]{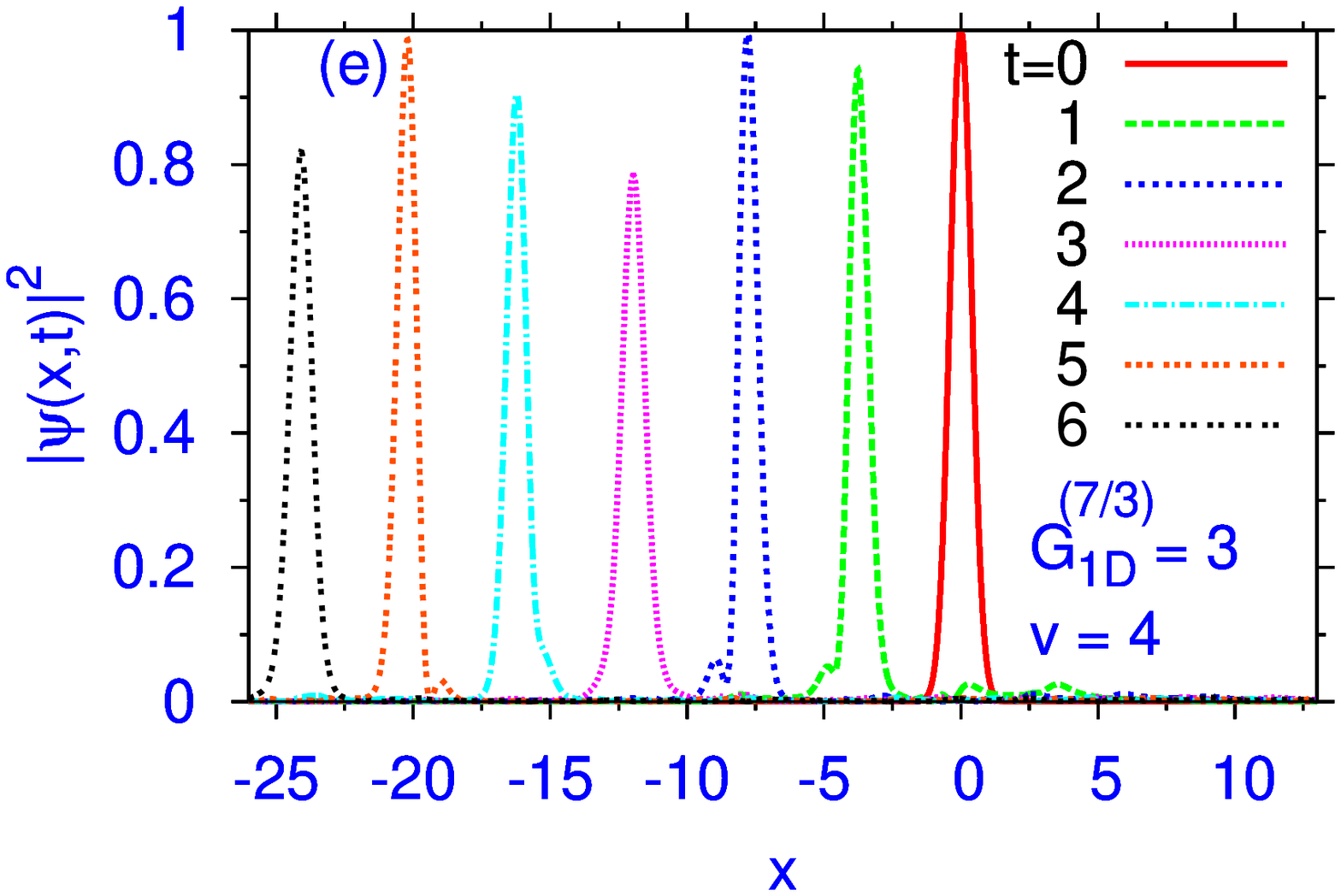} \includegraphics[width=.48%
\linewidth]{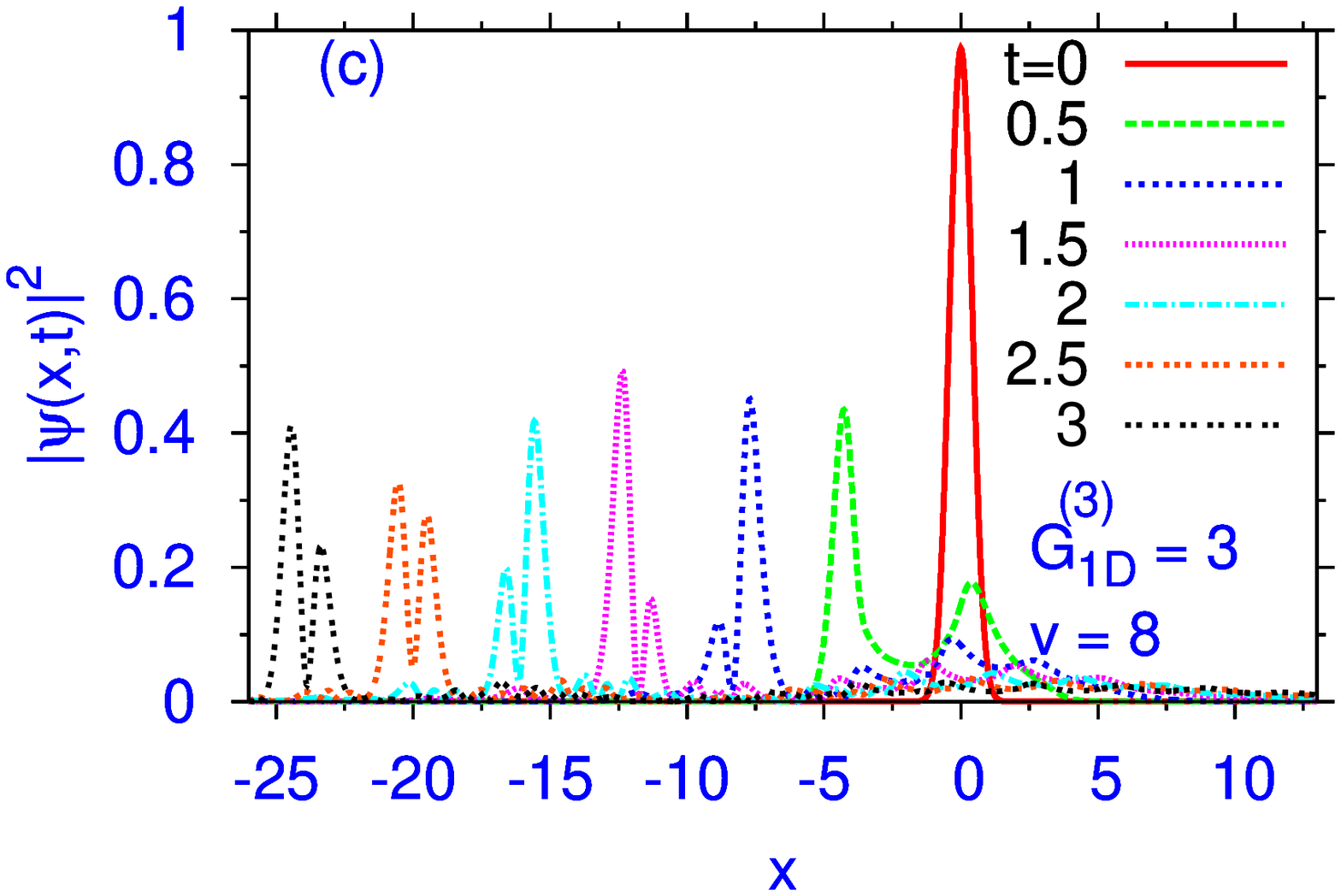} \includegraphics[width=.48\linewidth]{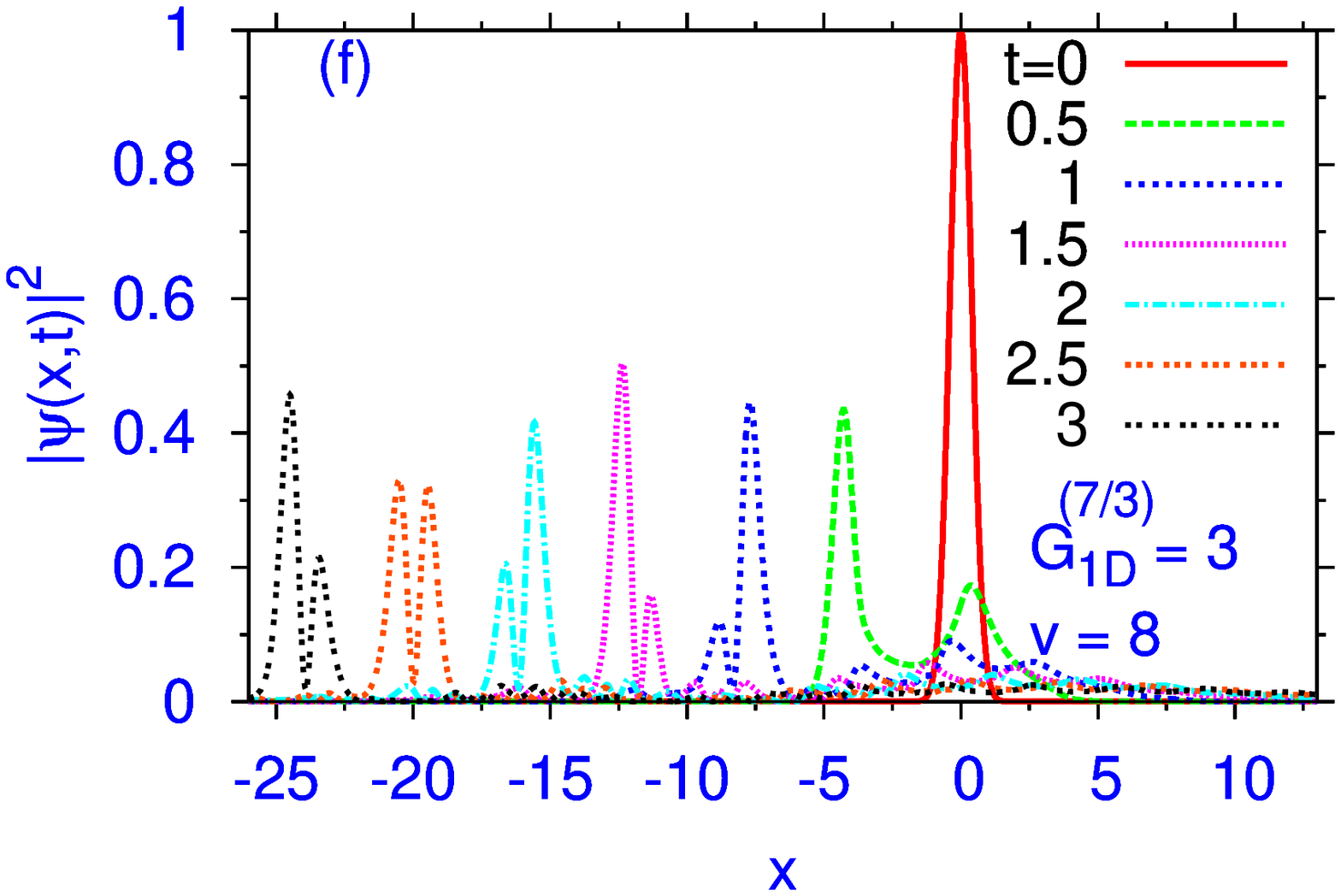}
\end{center}
\par
\caption{(Color online) Dragging GSs with $G_{\mathrm{1D}}^{(3)}=G_{\mathrm{%
1D}}^{(7/3)}=3$ by the OL potential $-5\cos [2(x+vt)]$, which was suddenly
set in motion at constant velocities $-v=-2,-4,-8$. The upper (BEC) and
lower (BCS superfluid) panels display a set of snapshots of the density,
taken at consecutive moments of time.}
\label{fig8}
\end{figure}

A problem of considerable interest is a possibility of controllable transfer
of solitons by a moving OL. We have studied this possibility by means of
direct simulations of Eqs. (\ref{1Dequation}) and (\ref{bosonic}), replacing
the static OL by $V(x)=-V_{0}\cos \left( 2(x+vt)\right) $, which moves at
constant velocity $-v$. Typical results for the dragging of GSs taken from
the first bandgap are displayed in Fig. \ref{fig8}, both for the bosonic
model with the ordinary cubic nonlinearity, and for the model of the BCS
superfluid.

As one might expect, the soliton is dragged in a relatively stable fashion
at smaller velocities, but gets destroyed if $v$ is too large. The quality
of the transportation may be characterized by the relative loss of the
number of atoms in the soliton at the end of the simulation; these data are
collected in Table 1 (recall that the initial norm of each soliton is $1$,
as per Eq. (\ref{N})). It is difficult to exactly identify a critical value
of $v$ at which the moving lattice destroys the GS; rather, the transition
from the stable dragging (at $v=2$ and $4$) to the destruction (at $v=8$) is
gradual. The results for the bosonic model with the quintic nonlinearity
(not shown here) are generally similar, although in that case the GS appears
to be more fragile, as its destruction commences at velocity $v$ somewhat
smaller than in the two models represented in Fig. \ref{fig8}. The above
physical estimates suggest that typical length and time units in the present
setting are $\sim \,1$ $\mu $m and $1$ ms, from which we infer that the
stable transportation of the GSs is possible for velocities $<1$ mm/s.

\begin{tabular}{|l|l|l|l|}
\hline
$v$ & $2$ & $4$ & $8$ \\ \hline
BEC & $1.5\%$ & $7\%$ & $30\%$ \\ \hline
BCS superfluid & $2\%$ & $8\%$ & $28\%$ \\ \hline
\end{tabular}

Table 1. The relative loss of atoms suffered by the one-dimensional gap
solitons at the end of dragging shown in Fig. \ref{fig8}.

\section{Two-dimensional solitons}

\subsection{Variational approximation}

Stationary solutions to Eq. (\ref{2Dequation}) are looked for as $\psi
(x,y,t)=e^{-i\mu t}\phi (x,y)$, with real function $\phi (x)$ obeying
equation%
\begin{equation*}
\mu \phi +(1/2)\left( \partial _{x}^{2}+\partial _{y}^{2}\right) \phi -G_{%
\mathrm{2D}}^{(7/3)}\phi ^{7/3}
\end{equation*}%
\begin{equation}
+V_{0}\left[ \cos \left( 2x\right) +\cos \left( 2y\right) \right] \phi =0,
\label{phi2D}
\end{equation}%
cf. Eq. (\ref{phi1D}) (the small cubic term is dropped here). Together with
normalization condition (\ref{N2D}), Eq. (\ref{phi2D}) can be derived from
the respective Lagrangian,
\begin{eqnarray}
L &=&\int \int dxdy\left\{ \mu \phi ^{2}-\frac{1}{2}\left( \nabla \phi
\right) ^{2}-\frac{3}{5}G_{\mathrm{2D}}^{(7/3)}\phi ^{10/3}\right.  \notag \\
&&\left. +V_{0}\left[ \cos (2x)+\cos (2y)\right] \phi ^{2}\right\} -\mu ,
\label{L2D}
\end{eqnarray}%
cf. Eq. (\ref{L}). To apply the VA, we adopt the 2D Gaussian ansatz, cf. its
1D counterpart (\ref{Gauss}), as
\begin{equation}
\phi (x,y)=\frac{1}{W}\sqrt{\frac{M}{\pi }}\exp \left( -\frac{x^{2}+y^{2}}{%
2W^{2}}\right) ,  \label{Gauss2D}
\end{equation}%
where $M$ is the 2D norm, that will be actually fixed as $M=1$, in
accordance with Eq. (\ref{N2D}) (for the time being, $M$ is one of the
variational parameters, together with $\mu $ and width $W$, like in the 1D
setting considered above). An anisotropic ansatz for 2D solitons, with
different widths in the $x$- and $y$-directions, was considered too. We do
not present it here, as numerical solutions of the variational equations
have revealed only isotropic solitons.

The substitution of ansatz (\ref{Gauss2D}) in Lagrangian (\ref{L2D}) yields
the effective Lagrangian, cf. Eq. (\ref{LGauss})],%
\begin{eqnarray}
L_{\mathrm{eff}} &=&\mu \left( M-1\right) -\frac{M}{2W^{2}}  \notag \\
&&-\frac{9G_{\mathrm{2D}}^{(7/3)}M^{5/3}}{25\pi ^{2/3}W^{4/3}}%
+2V_{0}Me^{-W^{2}}.  \label{Leff}
\end{eqnarray}%
The first variational equation, $\partial L_{\mathrm{eff}}/\partial \mu =0$,
yields, as expected, $M=1$. Then, equations $\partial L_{\mathrm{eff}%
}/\partial W=\partial L_{\mathrm{eff}}/\partial M=0$ take the following
form:
\begin{equation}
\frac{1}{2W^{4}}+\frac{6G_{\mathrm{2D}}^{(7/3)}}{25\pi ^{2/3}}\frac{1}{%
W^{10/3}}=2V_{0}e^{-W^{2}},  \label{W1}
\end{equation}%
\begin{equation}
\mu =\frac{1}{2W^{2}}-2V_{0}e^{-W^{2}}+\frac{3G_{\mathrm{2D}}^{(7/3)}}{5\pi
^{2/3}}\frac{1}{W^{4/3}}.  \label{muVA}
\end{equation}

In the linear limit, $G_{\mathrm{2D}}^{(7/3)}=0$, Eq. (\ref{W1}) coincides
with its 1D counterpart, Eq. (\ref{artifact}). Accordingly, physical
solutions ($W^{2}>0$) exist, in the linear limit, for $V_{0}>\left(
V_{0}\right) _{\min }=e^{2}/16\approx \allowbreak 0.462$, see Eq. (\ref{min}%
). Although the linear Schr\"{o}dinger equation with a periodic potential
cannot have localized solutions (in any dimension), the result obtained in
the linear limit makes sense, similar to the situation in the 1D model:
after taking $W$ as the smaller root of Eq. (\ref{artifact}), and then the
respective value of $\mu $ from Eq. (\ref{muVA}), one will obtain a value of
$\mu $ at which the family of the GS solutions emerges. In this way, the VA
predicts the left edge of the first finite bandgap in the 2D spectrum (a
similar result was obtained in \cite{GMM}, and for the quasi-1D OL potential
in the 2D setting -- also in work \cite{Thawatchai}, where the VA could
accurately predict the edge of the semi-infinite gap). In particular, the VA
gives $\left[ \left( \mu _{1}\right) _{\mathrm{left}}^{\mathrm{(var)}}\right]
_{\mathrm{2D}}\approx -4.26$ for $V_{0}=4$, whereas the numerical solution
of linearized equation (\ref{phi2D}) yields$\left[ \left( \mu _{1}\right) _{%
\mathrm{left}}^{\mathrm{(num)}}\right] _{\mathrm{2D}}\approx -4.25$ in this
case, cf. result (\ref{var}) and its numerical counterpart, obtained in the
1D case.

\subsection{Numerical results}

Numerical solutions for 2D solitons were obtained by running simulations of
Eq. (\ref{2Dequation}) until the solutions would settle down to stationary
real states, as done above for 1D equation (\ref{1Dequation}); recall this
procedure produces only stable solutions. The results were compared to the
predictions of the VA. For the purpose of the comparison with BEC, we also
include findings for the 2D GPE with the usual cubic nonlinearity, whose
stationary form is%
\begin{equation*}
\mu \phi +(1/2)\left( \phi _{xx}+\phi _{yy}\right) -G_{\mathrm{2D}%
}^{(3)}\phi ^{3}
\end{equation*}%
\begin{equation}
+V_{0}\left[ \cos (2x)+\cos (2y)\right] \phi =0,~  \label{2Dbosonic}
\end{equation}%
cf. one-dimensional equation (\ref{bosonic}). Solutions to this equation are
normalized as in Eq. (\ref{N2D}), hence the respective nonlinearity
coefficient is \cite{Luca} $G_{\mathrm{2D}}^{(3)}=2\sqrt{2\pi }\left(
a_{s}/a_{\mathrm{ho}}\right) N$, where $a_{\mathrm{ho}}$ characterizes the
tight confinement in direction $z$.

\begin{figure}[tbp]
\includegraphics[width=\linewidth,clip]{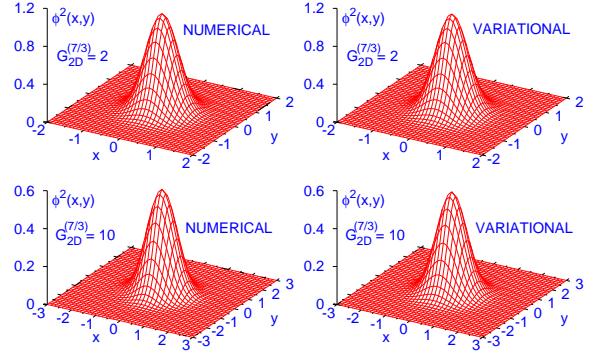}
\caption{(Color online) Examples of stable 2D GSs, obtained in the numerical
and variational forms from Eq. (\protect\ref{2Dequation}) for $V_{0}=4$. The
smaller and larger values of $G_{\mathrm{2D}}$ correspond to the first and
second bandgaps, respectively, see Fig. \protect\ref{fig11}.}
\label{fig9}
\end{figure}

Typical examples of the 2D GSs belonging to the first and second bandgaps of
the 2D spectrum, which are displayed in Fig. \ref{fig9} (and many other
examples not shown here) demonstrate that the VA provides a good fit to the
numerically found shapes, although, of course, simple Gaussian ansatz (\ref%
{Gauss2D}) does not capture very weak tails attached to the central body of
the solitons, that become (barely) visible in 2D numerical solutions taken
near the edge of the bandgap. In the bosonic model with the cubic
nonlinearity, the variational and numerically found soliton shapes are very
close too (not shown here).

To quantify the accuracy provided by the VA, we calculated the corresponding
relative error,
\begin{equation*}
\mathrm{err}=\frac{\int \int dxdy\left\vert |\phi _{\mathrm{num}%
}(x,y)|^{2}-|\phi _{\mathrm{var}}(x,y)|^{2}\right\vert }{\int \int dxdy|\phi
_{\mathrm{num}}(x,y)|^{2}}.
\end{equation*}%
For $G_{\mathrm{2D}}^{7/3}=2$ and $G_{\mathrm{2D}}^{7/3}=10$, which are the
cases displayed in Fig. \ref{fig9}, we have found $\mathrm{err}=0.020$, and $%
0.079$, respectively, which illustrates the accuracy provided by the VA for
2D solitons. With the increase of nonlinearity, the shape of the soliton
slightly deviates from the Gaussian, which gives rise to a larger value of 
error.

In addition to Eq. (\ref{2Dequation}) with the square-shaped OL, we have
also considered the same equation with the radial-lattice potential (the
small cubic term is dropped here),%
\begin{equation}
i\psi _{t}=-\frac{1}{2}\nabla ^{2}\psi -G_{\mathrm{2D}}^{(7/3)}|\psi
|^{4/3}\psi +V_{0}\cos \left( 2r\right) \psi =0.  \label{radial}
\end{equation}%
Unlike previously studied 2D models with the repulsive cubic nonlinearity
and radial potential of the Bessel type \cite{Barcelona}, the potential in
Eq. (\ref{radial}) does not vanish at $r\rightarrow \infty $, hence radial
GSs can be produced by this equation \cite{BBBradial}.
The shape of these solitons (not shown here) is quite similar to that of the
GSs supported, at the same values of $G_{\mathrm{2D}}^{(7/3)}$, by the
square lattice, which suggests that the square OL gives rise to nearly
isotropic solitons.

\begin{figure}[tbp]
\begin{center}
{\includegraphics[width=1.0\linewidth,clip]{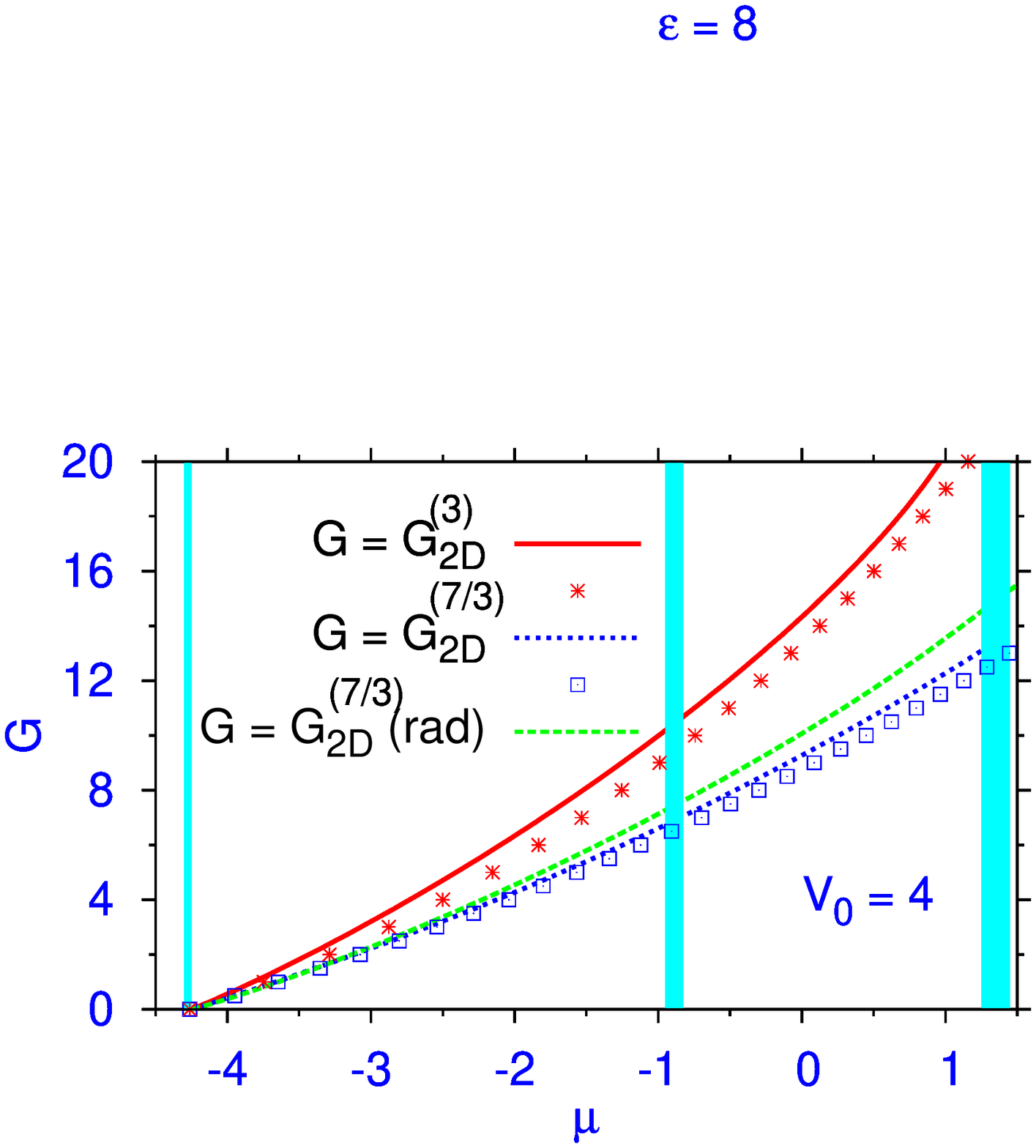}}
\end{center}
\caption{(Color online) The numerical (continuous lines) and variational
(strings of symbols) nonlinearity-versus-chemical-potential plots for 2D GSs
obtained from Eqs. (\protect\ref{2Dequation}) and (\protect\ref{2Dbosonic}).
The curve labeled \textquotedblleft $G=G_{\mathrm{3D}}^{(7/3)}$(rad)"
additionally displays numerical results for radial GSs generated by Eq. (%
\protect\ref{radial}) with the axisymmetric potential.}
\label{fig11}
\end{figure}

Bearing in mind normalization condition (\ref{N2D}), families of fundamental
two-dimensional GSs are characterized by dependence $G_{\mathrm{2D}%
}^{(7/3)}(\mu )$ (and its counterpart, $G_{\mathrm{2D}}^{(3)}(\mu )$, for
bosonic Eq. (\ref{2Dbosonic})), similar to the 1D model. These dependences,
as predicted by the VA and found from the numerical solutions, are displayed
in Fig. \ref{fig11} (the VA for Eq. (\ref{2Dbosonic}) was also based on
ansatz (\ref{Gauss2D})). Numerical results for the family of the radial GSs
obtained from Eq. (\ref{radial}) are included too.

A conclusion is that the VA provides good accuracy in describing the GS
families in both models with the square-lattice potential, although, as in
the 1D case, the approximation formally predicts that the families continue
across the narrow Bloch band separating the first and second bandgaps, where
solitons cannot exist. Another noteworthy similarity to the 1D case is that
the VA yields a higher accuracy for \ the nonlinearity of power $7/3$ than
for its cubic counterpart.

The family of 2D GSs was also constructed taking into regard the additional
cubic term in Eq. (\ref{2Dequation}). It was found that, as well as in the
1D case, this term, if taken with physically relevant values of the
coefficient in front of it (an estimate for which is given below), produced
a negligible effect.

\subsection{Physical estimates for the two-dimensional solitons}

As said above, underlying conditions (\ref{de Broglie}) and (\ref%
{semiclassical}) definitely hold for the 2D GSs generated by Eq. (\ref%
{2Dequation}). Proceeding to the estimate for the number of atoms in 2D
solitons in the BCS superfluid, it is relevant to mention that no two- (or
three-) dimensional matter-wave solitons of any type, regular or gap-mode,
have been created, thus far, even in BEC, therefore exploring new
possibilities for the creation of multidimensional matter-wave solitons may
be quite relevant to the experiment.

Figure \ref{fig11} demonstrates that achievable values of the normalized
nonlinearity strength for 2D GSs are $\left( G_{\mathrm{2D}}^{(7/3)}\right)
_{\max }\simeq 15$. Undoing rescalings (\ref{scaling2D}) and (\ref{g2D}), we
conclude that the corresponding number of atoms in the soliton is in the
range of $\left( N_{\mathrm{2Dsoliton}}\right) _{\max }\sim 10^{3}$, the
respective largest quantum number of the filled states in the transverse
trapping potential being $n\sim 10$. This number of atoms is sufficient to
justify the consideration of solitons in terms of the BCS superfluid, and to
make experimental observation of the solitons possible.

As for the coefficient in front of the additional cubic term in Eq. (\ref%
{2Dequation}), the estimate making use of expression (\ref{g2D})
demonstrates that it may achieve values $\left\vert g_{\mathrm{2D}%
}\right\vert \sim 0.1$, which are higher than $\left\vert g_{\mathrm{1D}%
}\right\vert \sim 10^{-3}$ in the 1D geometry (see above), but still quite
small in comparison with $G_{\mathrm{2D}}^{(7/3)}$.

\section{Conclusion}

The objective of this work was to predict the existence of quasi-1D\ and
quasi-2D solitons in the BCS superfluid formed by a gas of fermion atoms,
with weak attraction between atoms with opposite orientations of their
spins. We considered the experimentally relevant configuration of the gas
trapped under the combined action of the 1D or 2D optical lattice (OL) and
tight 2D or 1D trap applied in the transverse direction(s). The analysis was
based on the 3D equation for the wave function derived from the Lagrangian
that included the energy density of the weakly coupled BCS superfluid.
Then, the equation was reduced to effective 1D and 2D equations, taking into
regard the tight transverse confinement. A characteristic feature of the
equations is the self-repulsive nonlinear term of power $7/3$; it was also
shown that the next-order correction to the energy density of the weakly
coupled superfluid gives rise to a small additional cubic term in the
equations. Families of stable one- and two-dimensional GSs (gap solitons)
were found, by means of the VA (variational approximation) and in the
numerical form, in the first two finite bandgaps of the OL-induced spectrum.
The VA, based on the Gaussian ansatz, provides good accuracy in the
description of the families of 1D and 2D solitons, except for very close to
edges of the bandgaps, where the GS changes its shape from tightly- to a
loosely-bound one, with undulating tails. In the linear limit, the VA
accurately predicts borders (i.e., narrow Bloch bands) separating different
gaps in the spectrum. The comparison with the predictions for 1D and 2D
bosonic GSs, supported by the cubic (or quintic) repulsive nonlinearity, has
demonstrated that the VA provides a higher accuracy in predicting the
solitons in the present model with the weaker nonlinearity, of power $7/3$.
Even and odd stable bound states of one-dimensional GSs were found too. They
may be interpreted as pairs of fundamental solitons placed in adjacent local
wells of the lattice. In the 2D case, additionally found solutions are
radial GSs supported by the radial-lattice potential.

The quasi-1D and quasi-2D gap solitons, confined by the moderately strong
transverse potential, may contain the number of fermion atoms in the ranges
of $10^{3}-10^{4}$ and $10^{3}$, respectively. The possibility of dragging
GSs by a moving OL was studied in the 1D setting. It was demonstrated that
there is a smooth transition from a regime of stable motion of the soliton
to its destruction, as the OL's velocity increases beyond $\sim 1$ mm/s.

Experimental creation of the quasi-1D and quasi-2D gap solitons in the BCS
superfluid seems quite feasible. As concerns further developments of the
theory, issues of straightforward interest are 3D solitons, as well as
vortex solitons in two dimensions.

We appreciate valuable discussions with L. Salasnich. The work of S.K.A. was
supported in a part by FAPESP and CNPq (Brazil). The work of B.A.M. was
partly supported by the Israel Science Foundation through the
Center-of-Excellence grant No. 8006/03, and by the German-Israel Foundation
through Grant No. 149/2006. B.A.M. appreciates hospitality of the Institute
of Theoretical Physics at UNESP (S\~{a}o Paulo State University) and
financial assistance from FAPESP.

\section*{Appendix: one-dimensional subfundamental solitons}

The ordinary one-dimensional GPE with the repulsive cubic nonlinearity and
OL potential, which corresponds to the first line in Eq. (\ref{bosonic}),
gives rise to antisymmetric SF (subfundamental)\textit{\ }solitons, which
were found in the second bandgap of the OL-induced spectrum in work \cite%
{Thawatchai1D}. They are called so because, if one uses the notation with
fixed $G_{\mathrm{3D}}^{(3)}$ and arbitrary norm, these solitons have the
norm which is lower than in the fundamental bosonic GSs, for the same
chemical potential. A characteristic feature of the SF solitons is that two
maxima of the density, $|\phi (x)|^{2}$, are located inside a single cell of
the OL potential (i.e., this antisymmetric soliton as a whole is essentially
confined to the single cell). The SF solitons are different from
antisymmetric bound states of two fundamental solitons, which feature two
maxima of $\left\vert \phi (x)\right\vert ^{2}$ located in different
potential wells, see Fig. \ref{fig6}(a).

In the 1D model derived in this work, i.e., Eq. (\ref{1Dequation}), families
of SF solitons can also be found in the second bandgap. They are unstable,
although their instability is weak, as, otherwise, the numerical method
described above, which is based on the direct integration of Eq. (\ref%
{1Dequation}), would not reveal them. Similar to the situation in the GPE,
the instability does not completely destroy the SF solitons, but rather
converts them into fundamental GSs belonging to the first finite bandgap.

The SF solitons, as \emph{odd} solutions to Eq. (\ref{phi1D}), may be
represented by the VA based on the accordingly modified Gaussian ansatz,
\begin{equation}
\phi (x)=\pi ^{-1/4}\frac{\sqrt{2M}}{W^{3/2}}x\exp \left( -\frac{x^{2}}{%
2W^{2}}\right)   \label{Gauss1}
\end{equation}%
(its norm is $1$ for $M=1$). Inserting this ansatz in Lagrangian (\ref{L})
yields
\begin{eqnarray}
L &=&\mu \left( M-1\right) -\frac{3M}{4W^{2}}+{V}_{0}M(1-2W^{2})e^{-W^{2}}
\notag \\
&-&\frac{2^{5/3}\Gamma \left( 13/6\right) }{\left( 5/3\right) ^{19/6}\pi
^{5/6}}G_{\mathrm{1D}}^{(7/3)}\frac{M^{5/3}}{W^{2/3}}~,  \label{LGauss1}
\end{eqnarray}%
cf. Eq. (\ref{LGauss}). The first variational equation following from Eq. (%
\ref{LGauss1}), $\partial L/\partial \mu =0$, gives $M=1$, as before. The
remaining equations, $\partial L/\partial W=0$ and $\partial L/\partial M=0$%
, take the following form, cf. Eqs. (\ref{WGauss}) and (\ref{muGauss}):
\begin{eqnarray}
1 &+&\frac{2^{11/3}\Gamma \left( 13/6\right) }{9\left( 5/3\right) ^{19/6}\pi
^{5/6}}G_{\mathrm{1D}}^{(7/3)}W^{4/3}  \notag \\
&=&\frac{4}{3}V_{0}W^{4}(3-2W^{2})e^{-W^{2}},  \label{WGauss1}
\end{eqnarray}%
\begin{eqnarray}
\mu  &=&\frac{3}{4W^{2}}+\frac{2^{5/3}\Gamma (13/6)}{\left( 5/3\right)
^{13/6}\pi ^{5/6}}\frac{G_{\mathrm{1D}}^{(7/3)}}{W^{2/3}}  \notag \\
&-&{V_{0}}(1-2W^{2})e^{-W^{2}}.  \label{muGauss1}
\end{eqnarray}

Note that setting $G_{\mathrm{1D}}^{(7/3)}=0$ in Eq. (\ref{WGauss1}) yields
an equation for $W^{2}$ which has two physical roots if $V_{0}$ exceeds a
minimum value, $\left( V_{0}\right) _{\min }\approx 0.37$, cf. expression (%
\ref{min}) for the minimum value of $V_{0}$ in the VA based on ansatz (\ref%
{Gauss}). The smaller root, if substituted in Eq. (\ref{muGauss1}) with $G_{%
\mathrm{1D}}^{(7/3)}=0$, yields $\left( \mu _{2}\right) _{\mathrm{left}}^{%
\mathrm{(var)}}\equiv \mu \left( G_{\mathrm{1D}}^{(7/3)}=0\right) $, which,
as shown by the dotted curve in Fig. \ref{fig1}, accurately predicts the
lower/left edge of the second bandgap. In particular, this approximation
predicts $\left( \mu _{1}\right) _{\mathrm{left}}^{\mathrm{(var)}}\approx
1.02$ for $V_{0}=5$, while the respective value obtained from the numerical
solution of the Mathieu equation is $\mu _{12}^{\mathrm{(num)}%
}(V_{0}=5)\approx 1.03$, cf. variational prediction (\ref{var}) for the
border between the semi-infinite and first finite gaps.

\begin{figure}[tbp]
\begin{center}
{\includegraphics[width=1.0\linewidth]{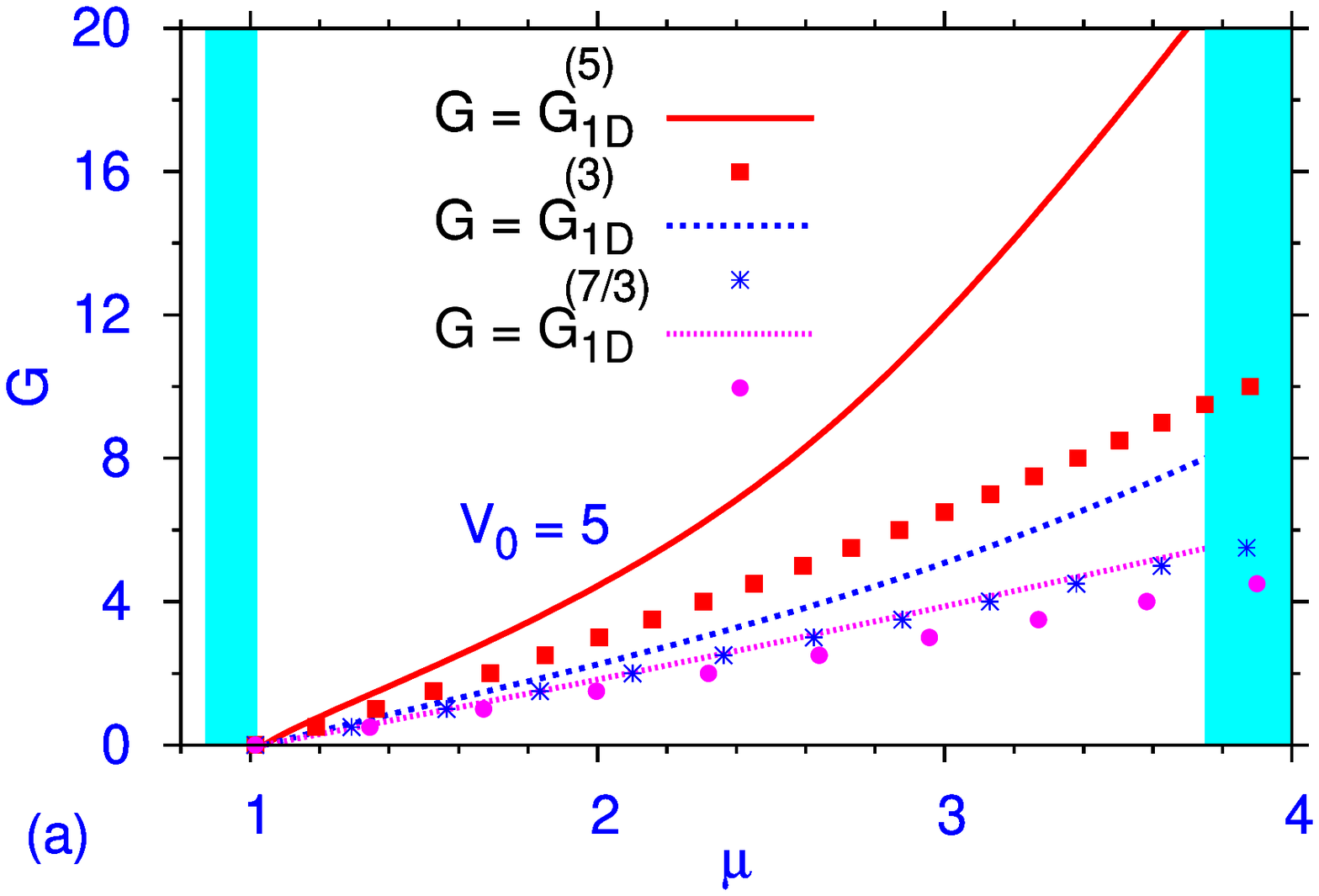}} {%
\includegraphics[width=1.0\linewidth,clip]{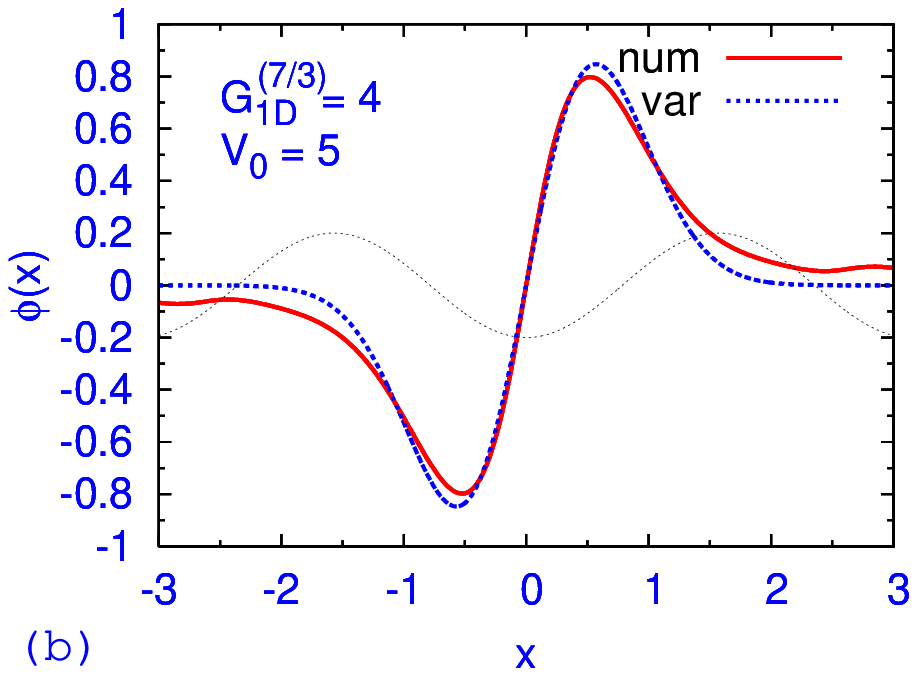}}
\end{center}
\caption{(Color online) (a) The same as in Fig. \protect\ref{fig2}, but for
1D subfundamental solitons, found in the second finite bandgap. (b) A
typical example of the subfundamental soliton found from the numerical
solution of Eq. (\protect\ref{phi1D}), and its variational counterpart.}
\label{fig12}
\end{figure}

The comparison of the dependence $G_{\mathrm{1D}}^{(7/3)}(\mu )$ for the SF
solitons in the second bandgap, as predicted by Eqs. (\ref{WGauss1}) and (%
\ref{muGauss1}), versus its numerically found counterpart is presented in
Fig. \ref{fig12}(a). Again, for the purpose of comparison, this figure
additionally includes dependences between the nonlinearity strength and
chemical potential of SF solitons, as obtained in the numerical form and by
means of the VA (that was also based on ansatz (\ref{Gauss1})) for two
bosonic equations (\ref{bosonic}). As seen from this figure and Fig. \ref%
{fig2}, the accuracy of the variational prediction for the SF soliton
families is worse than it was for the fundamental GSs. Nevertheless, the
prediction is still acceptable for the SF solitons in the BCS-superfluid
model. In particular, the shape of the SF solitons in this model is
approximated by the VA quite accurately, as shown in Fig. \ref{fig12}(b). On
the other hand, it is evident from Fig. \ref{fig12}(a) that the discrepancy
between the VA and numerical results for the SF solitons found in the
bosonic models is much larger, which strongly confirms the above inference
that the VA works essentially better in the model of the BCS superfluid than
in the BEC models, i.e., for the nonlinearity of power $7/3$ than for the
cubic and quintic nonlinear terms.

\end{document}